\RequirePackage{fix-cm}
\documentclass[twocolumn]{svjour3}          
\smartqed  
\usepackage{amssymb}
\usepackage{amstext}
\usepackage{amsmath}
\usepackage{multicol, multirow}
\usepackage{graphicx}
\usepackage{dsfont}
\usepackage{ctable}
\usepackage{pifont}
\usepackage[percent]{overpic}
\usepackage[T1]{fontenc}
\usepackage{url}

\usepackage{afterpage}

\newcommand{\norm}[1]{\lvert\lvert #1 \rvert\rvert}

\DeclareMathOperator*{\argmin}{arg\,min}

\begin{document}\sloppy

\title{Multi-view data capture for dynamic object reconstruction using handheld augmented reality mobiles}

\author{
Matteo Bortolon \and
Luca Bazzanella \and
Fabio Poiesi
}


\institute{M. Bortolon and L. Bazzanella \at
            Universit\'a degli Studi di Trento\\
            Trento, Italy\\
            \email{matteo.bortolon@studenti.unitn.it}\\
            \email{luca.bazzanella-2@studenti.unitn.it}\\
           \and
           F. Poiesi (corresponding author) \at
           Technologies of Vision, Fondazione Bruno Kessler\\
           Trento, Italy\\
           \email{poiesi@fbk.eu}
}

\date{Received: date / Accepted: date}

\maketitle

\begin{abstract}
We propose a system to capture nearly-synchronous frame streams from multiple and moving handheld mobiles that is suitable for dynamic object 3D reconstruction. 
Each mobile executes Simultaneous Localisation and Mapping on-board to estimate its pose, and uses a wireless communication channel to send or receive synchronisation triggers.
Our system can harvest frames and mobile poses in real time using a decentralised triggering strategy and a data-relay architecture that can be deployed either at the Edge or in the Cloud.
We show the effectiveness of our system by employing it for 3D skeleton and volumetric reconstructions. 
Our triggering strategy achieves equal performance to that of an NTP-based synchronisation approach, but offers higher flexibility, as it can be adjusted online based on application needs. 
We created a challenging new dataset, namely 4DM, that involves six handheld augmented reality mobiles recording an actor performing sports actions outdoors. 
We validate our system on 4DM, analyse its strengths and limitations, and compare its modules with alternative ones.
\end{abstract}

\section{Introduction}\label{sec:intro}

Dynamic objects recorded from multiple viewpoints can be digitally reconstructed to produce free-viewpoint videos (FVV), or 4D videos, which can then be experienced in Augmented or Virtual Reality \cite{Pages2018,Qiao2019,Mustafa2019b}.
Multi-view 3D/4D reconstruction and understanding of dynamic objects is challenging because it requires cameras to be calibrated in space and time \cite{Leo2008,Kim2012,Collet2015,Mustafa2017,Mustafa2019,Bortolon2020}. 
The 4D reconstruction task in uncontrolled scenarios for real-time applications (e.g.~telepresence) is particularly challenging as it requires accurate camera synchronisation, calibration and communication to operate seamlessly on commodity hardware \cite{Richardt2016,Mustafa2017}.
FVV is sensitive to frame-level synchronisation inaccuracies because frame misalignments may cause 3D reconstruction artifacts.
For controlled setups, frame-level synchronisation can be achieved using shutter-synchronised cameras \cite{Mustafa2017}. 
However employing this approach in uncontrolled environments with conventional mobiles is unrealistic.
Also, camera calibration must be guaranteed at frame level in order to triangulate 2D features in 3D over time when handheld mobiles are used.
Typically, with static cameras, an expert operator uses a checkerboard or a calibration wand to estimate intrinsic and extrinsic parameters prior to the data capture \cite{Collet2015}.
With moving cameras, an option can be to exploit (external) static cameras to compute the extrinsic parameters of the moving ones \cite{Kim2012}.
However, this approach is also unrealistic in uncontrolled environments as it requires an ad-hoc setup.
Lastly, to avoid frame transmission delays or drops due to packet buffering, bandwidth congestions should be carefully handled \cite{Bortolon2020}.

\begin{figure}[t]
  \centering
  \includegraphics[width=1\columnwidth]{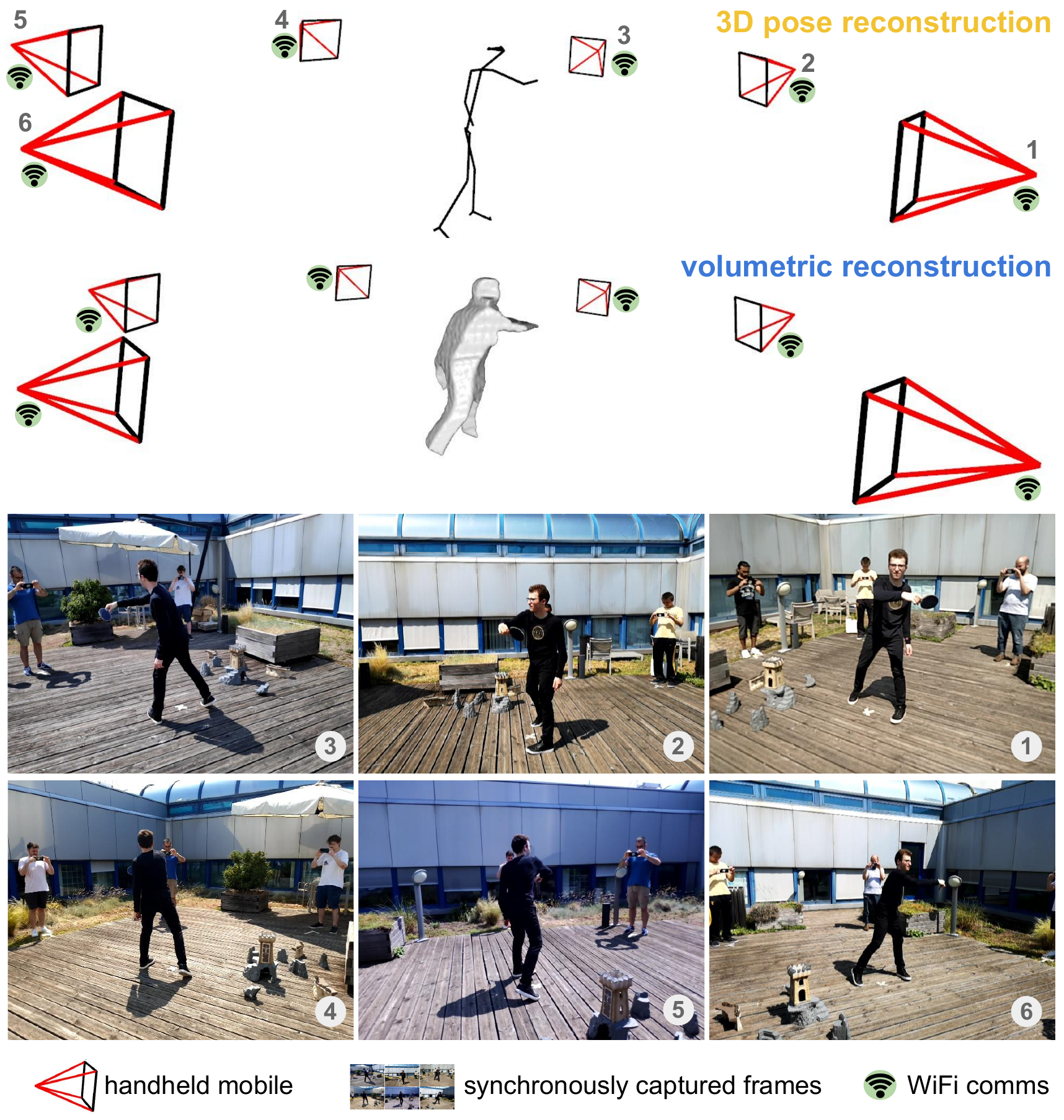}
  \caption{Six people recording an actor from different viewpoints with their augmented reality-enabled mobiles.
  Our system can synchronously capture mobile poses and camera frames from the mobiles, thus allowing applications such as 3D skeleton reconstruction or volumetric reconstruction.
  Mobiles communicate through conventional wireless connections (e.g.~802.11, 5G).}
  \label{fig:teaser}
\end{figure}

In this paper, we propose the first mobile-based system that is designed for multi-view dynamic object reconstruction in natural scenes using commodity hardware (Fig.~\ref{fig:teaser}).
Our setup employs multiple mobiles that can estimate their own pose on-board in real time via Simultaneous Localisation and Mapping (SLAM).
We use ARCore's SLAM implementation running on the Android mobiles \cite{arcore}.
A host mobile must be the initiator of a capture session.
A set of client mobiles can then join this session by connecting to the host.
Mobiles can capture nearly-synchronous frames and their respective poses using wireless communications (e.g.~802.11, 5G).
The host uses synchronisation triggers to inform the clients when to capture frames and their respective poses.
To mitigate inter-mobile communication delays, we use a delay compensation strategy to temporally align host and clients' frame captures through an artificial delay that is applied to the frame capturing after the arrival of the trigger.
Each mobile applies a different delay that is based on its own round trip time (RTT) with respect to the host and on the largest RTT measured by all the mobiles.
We experimentally show that our synchronisation triggering strategy equals the performance of an NTP-based approach \cite{Mills1991}, but offers the advantage of being more flexible, as the trigger frequency can be changed online based on application needs.
We collected a new outdoor dataset, namely 4DM (4DMobile), which involves an actor performing sports actions that we recorded using six handheld augmented reality mobiles.
To show that our system is effective for dynamic object reconstruction, we implemented and evaluated two challenging applications, namely temporal 3D skeleton and volumetric reconstructions. 
In the former, we reconstructed the 3D skeleton of the actor by triangulating its 2D poses estimated on each mobile image plane.
In the latter, we reconstructed the volume of the person from their 2D silhouettes.
Fig.~\ref{fig:teaser} shows an example of our results.
The source code and the 4DM dataset are avalilable at \url{https://github.com/fabiopoiesi/4dm}.

In summary, the novelties of our work are:
\begin{itemize}
\item A mobile-based system to reconstruct dynamic objects in natural scenes that can be deployed using commodity hardware.
Unlike \cite{Kim2012}, our system does not require manual camera calibration;
\item A delay compensation strategy to allow nearly-synchronous data captures from multiple viewpoints for wirelessly-connected mobiles.
Unlike \cite{Pages2018}, which uses an ultra-bright LED flash and synchronises frames in post-processing, we do not use specialised hardware.
Unlike \cite{Ansari2019}, which operates on the camera sensor directly and exploits the mobile hardware image streaming, our synchronisation problem is more challenging because the captured frames have to be synchronised with their corresponding mobile poses and across mobiles, and because the SLAM process introduces delays that are non-deterministic.
SLAM runs on top of the Unity3D rendering engine, needed by the UI/UX to provide guidelines to the users during the frame capture, thus leading to irregular time intervals between consecutive frames;
\item A system that significantly extends our previous work \cite{Bortolon2020}: we re-designed the module in charge of managing frame streams in real time, making it more efficient, and implemented two new applications to assess the quality of the captured data;
\item A dataset (4DM) for 4D object reconstruction that we collected using six consumer handheld augmented reality mobiles in natural scenes. Unlike \cite{Kim2012}, 4DM is captured with consumer mobile cameras. Unlike \cite{Ballan2010}, the mobile poses are estimated in real time using SLAM. Unlike \cite{Sigal2009}, 4DM includes scenes captured with intentionally unstable handheld cameras.
\end{itemize}

The paper is organised as follows. 
Sec.~\ref{sec:related_work} presents related work about low-latency immersive computing and free-viewpoint video production. 
Sec.~\ref{sec:prob_form} formulates the problem we aim to tackle.
Sec.~\ref{sec:system_design} describes the design of our mobile-based system to capture near-synchronous frames and mobile poses.
Sec.~\ref{sec:application_and_results} describes the applications we implemented to validate our system.
Sec.~\ref{sec:results} presents the analysis of the experimental results.
Sec.~\ref{sec:concl} draws conclusions.

\section{Related work}\label{sec:related_work}

\noindent\textbf{Low-latency immersive computing:} 
To foster immersive interactions between multiple users in augmented spaces, low-latency computing and communications must be supported \cite{Yahyavi2013}. 
FVV is very sensitive to synchronisation issues and computations must be executed as close to the user as possible to reduce lag.

Synchronisation can be achieved using hardware- or software-based solutions.
Hardware-based solutions include timecode with or without genlock \cite{Kim2012}, and Wireless Precision Time Protocol \cite{Garg2018}.
Genlock is robust to drifts, but needs specialised network components.
Hardware-based solutions are not our target, as they require important modifications to the communication infrastructure.
Software-based solutions can be offline or online.
Offline solutions can use Spatio-Temporal Bundle Adjustment (ST-BA) to jointly compute the environment structure while temporally aligning frames \cite{Vo2016}.
However, ST-BA is a post-process with computational overhead, hence unsuitable for real-time applications.
Online solutions can use the Network Time Protocol (NTP) to instruct devices to capture frames periodically while attempting to compensate for delays \cite{Latimer2015,Wang2015,Ansari2019}. 
Cameras can share timers that are updated by a host camera \cite{Wang2015}. 
Software solutions, if combined with ad-hoc hardware control, can achieve state-of-the-art synchronisation accuracy \cite{Ansari2019}.
Although NTP approaches are simple to implement, they are unaware of situational context. 
Hence, the way in which clients are instructed to capture images in a session is totally disconnected from scene activity. 
These clients are unable to optimise acquisition rates either locally or globally, prohibiting optimisation techniques such as \cite{Poiesi2017}, that aim to save bandwidth and maximise output quality.
Our solution operates online and is aimed at decentralising synchronisation supervision, thus is more appropriate for resource-efficient, dynamic-scene capture.

\noindent\textbf{Free-viewpoint video production:} 
Free-viewpoint (volumetric or 4D) videos can be created either through the synchronised capturing of objects from different viewpoints \cite{Guillemaut2011,Mustafa2017}, or with Convolutional Neural Networks (CNN) \cite{Rematas2018} that estimate unseen content. 
The former strategy needs camera poses to be estimated/known for each frame, using approaches like SLAM \cite{Zou2013} or by having hardware-calibrated camera networks \cite{Mustafa2017}. 
Typically, estimated poses lead to less-accurate reconstructions \cite{Richardt2016} when compared to calibrated setups \cite{Mustafa2017}. 
Conversely, CNN-based strategies do not need camera poses, but instead need synthetic training data of the same type as the observed 3D objects, for example from video games \cite{Rematas2018}.
Traditional FVV (i.e.~non-CNN) approaches can be based on shape-from-silhouette \cite{Guillemaut2011}, shape-from-photoconsistency \cite{Slabaugh2001}, multi-view stereo \cite{Richardt2016} or deformable models (\cite{Huang2014}. 
Shape-from-silhouette aims to create 3D volumes (or visual hulls) from the intersections of visual cones formed by 2D outlines (silhouettes) of objects visible from multiple views. 
Shape-from-photoconsistency creates volumes by assigning intensity values to voxels (or volumetric pixels) based on pixel-colour consistencies across images. 
Multi-view stereo creates dense point clouds by merging the results of multiple depth maps computed from multiple views. 
Deformable model-based methods aim to fit known reference 3D models to visual observations, e.g.~2D silhouettes or 3D point clouds. 
All these methods need frame-level synchronised cameras.
\cite{Vo2016} describes a spatio-temporal bundle adjustment algorithm to jointly calibrate and synchronise cameras.
Because it is a computationally costly algorithm, it is desirable to initialise it with reliable camera poses and synchronised frames. 
Amongst these methods, multi-view stereo produces reconstructions that are geometrically more accurate than the other alternatives, albeit at a higher computational cost. Approaches like shape-from-silhouette and shape-from-photoconsistency are more suitable for online applications as they are fast, but outputs have less definition.
We implemented a proof-of-concept shape-from-silhouette method based on the frames and the mobile camera poses that are captured with our system.

\section{Problem formulation}\label{sec:prob_form}

Let $\mathcal{C} = \{ c_j \}_{j=1}^C$ be a set of $C$ handheld mobiles. 
Let $f_{i,c}$ be the $i^{th}$ frame captured by camera $c \in \mathcal{C}$. 
$f_{i,c}$ is an image composed of $N$ pixels $\mathcal{X} = \{x_n\}_{n=1}^N$, where $x_n \in \mathbb{R}^2$ is the $n^{th}$ pixel. 
Let $K_c \in \mathbb{R}^{3\times3}$ be the camera matrix (intrinsic parameters), $R_{i,c} \in \mathbb{R}^{3\times3}$ be the camera rotation and $t_{i,c} \in \mathbb{R}^3$ be the camera translation (extrinsic parameters) with respect to each camera's local reference. $R_{0,c} = \boldsymbol{I}_{3\times3}$ and $t_{0,c} = [0,0,0]^\textsf{T}$, where $\boldsymbol{I}_{3\times3}$ is a 3$\times$3-identity matrix and $\textsf{T}$ is the transpose operator.
Let $P_{i,c} = [R_{i,c} | t_{i,c}]$, $P_{i,c} \in \mathbb{R}^{3\times4}$, be the \emph{mobile pose}.
Each mobile is globally localised through a transformation $T_c \in \mathbb{R}^{3\times4}$ that maps $P_{i,c}$ to a global reference.
Let $D_{i,c} = \{f_{i,c}, P_{i,c}, \sigma_i\}$ be the data retrieved by $c$ upon the arrival of the \emph{synchronisation trigger} (or \emph{trigger}) $\sigma_i$. 
$\sigma_i$ is uniquely identified through an identifier.
Each $c$ aims to synchronously capture $D_{i,c}$ and transmit it in a timely manner to a central computing node. 
The goal of our system is to retrieve $D_{i,c} \forall c \in \mathcal{C}$ in real time and to use $D_{i,c}$ to generate 4D videos.

\section{Our mobile-based capture system}\label{sec:system_design}

\begin{figure}[t]
  \centering
  \includegraphics[width=1\columnwidth]{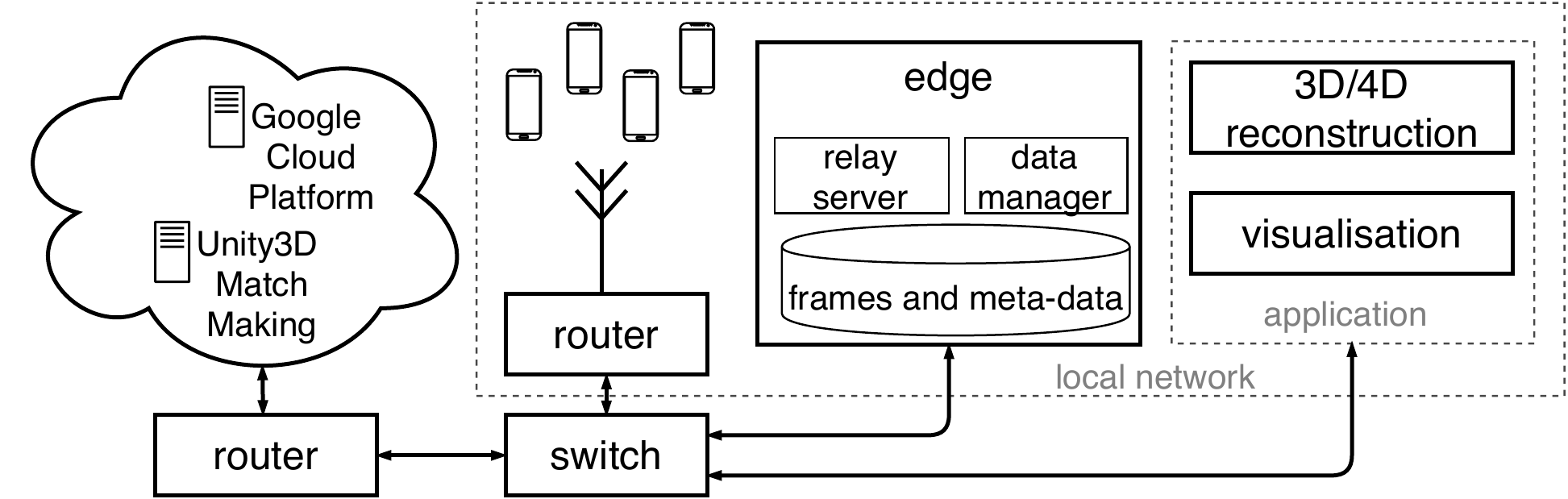}
  \caption{Block diagram of the proposed data-capture system, which involves a mobile host (the initiator of a capture session), a set of mobile clients (the participants), a Relay Server to handle synchronisation triggers and a Data Manager to process the data. 
  Mobiles communicate with Cloud services only to initiate the capture session.
  }
  \label{fig:block_diagram}
\end{figure}

Our data-capturing system mainly involves a mobile \emph{host} (the initiator of a capture session), a set of mobile \emph{clients} (the participants), a Relay Server to handle triggers and a Data Manager to process the data (Fig.~\ref{fig:block_diagram}). 
Mobiles aim to capture the data $D_{i,c}$ synchronously from different viewpoints. 
To decide when to capture $D_{i,c}$, the host generates triggers, rather than fixing the rate beforehand on a central node.
In this work we configured the host to send triggers at frequency $\phi$.
Note that a different policy for sending triggers can be implemented, e.g.~triggers can be generated dynamically at points that depend on the scene content.
When the host captures $D_{i,c}$, a counter is incremented, and its value is embedded in the trigger $\sigma_i$ and used as a unique identifier.
$D_{i,c}$ is then captured by each client after they receive $\sigma_i$.

\subsection{Session initialisation}

\subsubsection{Global localisation} 

Mobiles independently estimate the environment structure and their pose $P_{i,c}$ in real time through SLAM \cite{arcore}.
Once each mobile has mapped its surroundings, the host computes the features to estimate the transformation $T_c \forall c \in \mathcal{C}$. 
These features are sent to the Cloud, where they are processed to determine $T_c$ \cite{cloudanchors}. Mobiles receive $T_c$ and compute the mobile pose $P'_{i,c}$ = $T_c \circ P_{i,c}$ to globally localise themselves. $\circ$ is the composition operator. 
Without loss of generality, we keep $T_c \forall c \in \mathcal{C}$ fixed during the capture session.

\subsubsection{Latency estimation}\label{sec:latency_estimation}

For a given $\sigma_i$, if mobiles undergo delay, the dynamic object may be captured with a different pose from different viewpoints, thus affecting the quality of the 4D video.
The latency between host and clients can increase depending on the transmission distance, the network traffic, and the interference with other networks \cite{Soret2014}.
Therefore, latency must be analysed and compensated for \cite{Vo2016}.
We use a latency compensation strategy where the host measures the Round Trip Time (RTT) between itself and the clients, and communicates to each client the delay they must apply when capturing $D_{i,c}$.
The host builds a matrix $L \in \mathbb{R}^{A \times B}$, where the element $l_{a,b}$ is the $b^{th}$ measured RTT between the host and the $a^{th}$ client. $B$ is the number of RTT measurements and $A$=$C$-1 is the number of clients. 
The host computes the RTT for each client 
\begin{equation}
\bar{l}_a = \frac{1}{B} \sum_{b=1}^B l_{a,b}
\end{equation}
and then computes the maximum RTT
\begin{equation}
\ell = max(\{\bar{l}_1,...,\bar{l}_A\})
\end{equation}
The $a^{th}$ client will capture data with a delay of 
\begin{equation}
\Delta t_a = \frac{1}{2} ( \ell - \bar{l}_a )
\end{equation}
while the host will capture data with a delay of $\frac{1}{2} \ell$.

\subsection{Synchronisation relay}

To reduce the network traffic generated by trigger counts and to promote deployment (e.g.~802.11, 5G), we design a Relay Server to forward triggers received by the host to the clients.
Relay-based architectures have the disadvantage of increased latency compared to peer-to-peer ones \cite{Hu2016}, but we practically observed that they ease deployment, especially in networks that use Network Address Translation.
For timely delivery of the triggers, we employ a relay mechanism that is typically used for collaborative virtual environments \cite{Pretto2017} and teleoperation \cite{Yin2019}, namely MLAPI \cite{mlapi_relay}. 
MLAPI uses UDP to avoid delays caused by the flow-control system of TCP.
Our Relay Server is based on the concept of Multi-access Edge Computing and exploits the design of MLAPI. Its architecture and synchronisation protocols are detailed in \cite{Bortolon2020}.

\subsection{Data stream}\label{sec:data_management}

\begin{figure}[t]
  \centering
  \includegraphics[width=1\columnwidth]{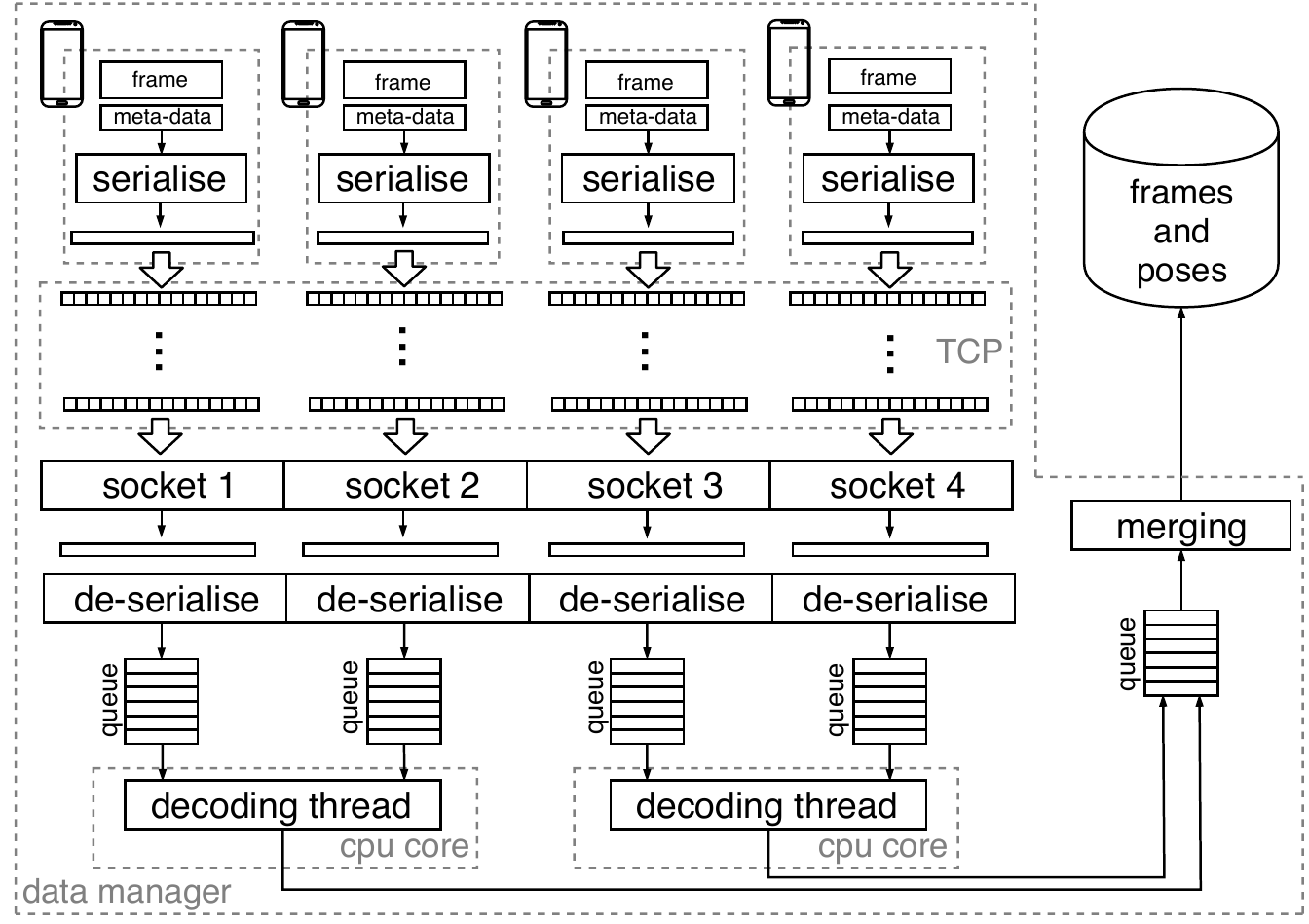}
  \caption{Data (i.e.~mobile poses and camera frames) is serialised on each mobile and transmitted to the data manager via TCP socket.
  Using a multi-thread-based application, the data manager de-serialises the data and checks for data consistency across what was received from the different mobiles via the merging operation.
  Captured data can be stored in a database for further processing.}
  \label{fig:data_manager}
\end{figure}

Throughput linearly increases with the number of mobiles until network congestion is reached.
We observed that four mobiles transmitting at 10fps on a conventional 2.4GHz WiFi suffice to reach maximum bandwidth.
Although HTTP-based applications are network-agnostic, they are unreliable with large throughputs.
When the maximum bandwidth capacity is reached, numerous HTTP-connections undergo time-outs due to delayed or lost TCP packets.
This leads to re-transmissions: more HTTP connections are created and more packets are dropped due to Socket buffer overflows.
Hence, triggers are not delivered (as they are UDP packets), leading to partial data captures.
To mitigate these problems, we design a Socket-based architecture that establishes a single connection between each mobile and the Data Manager (Fig.~\ref{fig:data_manager}).
This allows us to increase the percentage of successfully delivered packets.
Specifically, after $D_{i,c}$ is captured, the frame $f_{i,c}$ is encoded in JPEG, as it provides the best trade-off between compression and encoding/decoding speed.
$D_{i,c}$ is serialised \cite{protobuf}, split into packets and transmitted to the Data Manager via TCP.
Mobiles can eventually buffer packets if they detect network congestions.
Frames can also be video encoded to reduce the throughput, but this would require additional on-board processing. 
Moreover, to our knowledge, there are no video encoders that can embed metadata (like mobile poses) positionable with frame precision within a video stream.
The Data Manager receives the packets, merges them to restore the (original) serialised message and de-serialises the message.
De-serialisation can ingest up to 5282fps on a Xeon 3.1Ghz.
We associate each de-serialisation instance to a Socket. 
Frames are processed at 200fps by a verification module that checks their integrity.
We use a queue for each socket and perform verification using multiple threads.
Because frames are often disordered after verification, we re-organise them together with their meta-data (merging) before storing them in a database.

\section{Validation applications}\label{sec:application_and_results}

We evaluate our system through two applications that require synchronous frame captures, i.e.~the 3D skeleton reconstruction and the volumetric reconstruction af an actor over time.
In the former we estimate the actor's skeleton joints in 2D on each camera image plane and triangulate them in 3D.
In the latter, we volumetrically reconstruct the actor using a shape-from-silhouette-based approach \cite{Laurentini1994} because the black clothes hinder the extraction and matching of reliable and sufficient feature points.
We experienced this issue in our prior work \cite{Bortolon2020}.
Formally, let $\mathcal{O}$ be the actor. At each $i$ we process $D_{i,c} \forall c \in \mathcal{C}$ with the same trigger identity $\sigma_i$.

\subsection{3D skeleton reconstruction}\label{sec:3d_pose_estimation}

We extract the body pose of each person in $f_{i,c} \forall c \in \mathcal{C}$ \cite{Cao2017,wu2019detectron2}.
OpenPose \cite{Cao2017} can be used to estimate the 2D skeleton of each person using up to 25 joints per skeleton.
Because $\mathcal{O}$ is typically viewed in close-up on each mobile and in the centre of the frame, for each $f_{i,c}$ we score each detection as the product between the area of the bounding box enclosing the skeleton, and the inverse of the distance between the bounding box centre and the frame centre.
We select the skeleton that is inclosed in the bounding box with max score.
Let $H_{i,c} = \{h_{i,c}^m : h_{i,c}^m \in \mathbb{R}^2\}_{m=1}^M$ be the set of $M$ ordered skeleton joints (e.g.~$M$=25) of $\mathcal{O}$ estimated in $f_{i,c}$. $h_{i,c}^m$ is the position of the $m$th joint. 
We find the 3D joints by triangulating the corresponding 2D joints from camera pairs.
To compute reliable 3D joints, we use the camera pairs that have a predefined angle between their image plane normals.
Specifically, given a camera pair $c,c'\in \mathcal{C}$ such that $c \neq c'$, the angle between their image plane normals is defined as
\begin{equation}\label{eq:baseline_camera_pair}
\alpha_{i,c,c'} = \cos^{-1}\frac{\hat{n}_{i,c} \cdot \hat{n}_{i,c'}}{\norm{\hat{n}_{i,c}} \cdot \norm{\hat{n}_{i,c'}}},
\end{equation}
where $\hat{n}_{i,c}$ and $\hat{n}_{i,c'}$ are the image plane normals of camera $c$ and $c'$, respectively.
We deem $c,c'$ a valid pair if $\tau_\alpha \leq \alpha_{i,c,c'} \leq \tau_\alpha'$, where $\tau_\alpha$ and $\tau_\alpha'$ are parameters.
Let $\Gamma_i = \{\{c,c'\} : c,c' \in \mathcal{C}, c \neq c'\}$ be the set of valid camera pairs, $\forall \, \{c,c'\} \in \Gamma_i$ we compute the set of visible 3D joints.
We denote the $m^{th}$ 3D joint computed from $\{c,c'\}$ as 
\begin{equation}
\omega_{i,c,c'}^m = \mathrm{t} \left( h_{i,c}^m, h_{i,c'}^m, K_c P'_{i,c}, K_c P'_{i,c'} \right),
\end{equation}
where $\mathrm{t}(\cdot)$ is a triangulation algorithm, e.g.~the Direct Linear Transform algorithm (Sec.~12.2 in \cite{Hartley}).

Because each camera pair produces a 3D point for each joint, we typically have multiple 3D points for the same joint.
Let $\Omega_i^m = \{\omega_{i,c,c'}^m : \omega_{i,c,c'}^m \in \mathbb{R}^3, \{c,c'\} \in \Gamma_i \}$ be the set of 3D points of the $m^{th}$ joint triangulated from $\{c,c'\}$ and let $\Omega_i = \{\Omega_i^m\}_{m=1}^M$ be the set of all triangulated 3D points. 
We seek a 3D point for each joint to reconstruct the 3D skeleton.
Let $\mathcal{S}_i = \{s_i^m : s_i^m \in \mathbb{R}^3\}_{m=1}^M$ be the set of the skeleton's 3D joints. 
We design a function $g : \Omega_i \rightarrow \mathcal{S}_i$ that transforms a set of 3D points into a single 3D point for each joint as follows. 
A 3D point seen from multiple viewpoints can be reliably computed by minimising the re-projection error through Bundle Adjustment (BA) \cite{Lourakis2009} as
\begin{equation}\label{eq:bundle_adjustment}
\argmin_{\mathcal{A}_i, \mathcal{S}_i} \sum_{m=1}^M \sum_{c=1}^C d(Q(a_{i,c}, s_i^m), h_{i,c}^m)^2
\end{equation}
where $\mathcal{A}_i = \{a_{i,c} : c \in \mathcal{C} \}$ and $a_{i,c}$ is a vector whose elements are the intrinsic and extrinsic parameters of camera $c$ at frame $i$, i.e.~$K_c$, $P'_{i,c}$, respectively.
$Q(a_{i,c}, s_i^m)$ is the predicted projection of the joint $m$ on view $c$, and $d(x,y)$ is the Euclidean distance between the image points represented by the inhomogeneous vectors $x$ and $y$. 
The solution of Eq.~\ref{eq:bundle_adjustment} is a new set of camera parameters and 3D points. 
Without loss of generality we keep the camera parameters fixed and let BA determine only the new 3D points.
The non-linear minimisation of Eq.~\ref{eq:bundle_adjustment} requires initialisation. 
Snavely et al.~\cite{Snavely2007} use the camera parameters and the 3D points triangulated from an initial camera pair as initialisation, then BA is executed incrementally each time a new camera is added, using the new solution as initialisation for the next step (see Sec.~4.2 of \cite{Snavely2007}). 
This requires multiple iterations of BA at each $i$. 
To reduce the processing time, we execute BA globally, once for each $i$, using the centre of mass of each joint as initialisation. The $m^{th}$ centre of mass is
\begin{equation}\label{eq:ba_initialisation}
\mu_i^m = 
\begin{cases} 
\frac{1}{|\Omega_i^m|} \sum_{\{c,c'\} \in \Gamma_i} \omega_{i,c,c'}^m, & \mbox{if } \Omega_i^m \neq \emptyset \\ 
0, & \mbox{otherwise}
\end{cases}
\end{equation}
where $|\cdot|$ is the cardinality of a set. 
If $\mu_i^m = 0$, it will not be processed by BA. 
The solution of Eq.~\ref{eq:bundle_adjustment} provides $\mathcal{S}_i$.
We will empirically show in Sec.~\ref{sec:reprojection_error} that initialising with the centres of mass can reduce the computational time and that BA converges to the same solution as using the incremental approach of \cite{Snavely2007}.

\subsection{Volumetric reconstruction}
We extract the silhouette of each person in $f_{i,c} \forall c \in \mathcal{C}$ using instance-based segmentation \cite{He2017,Kirillov2020}.
We use PointRend \cite{Kirillov2020} as it provides superior performance in terms of silhouette segmentation to Mask R-CNN \cite{He2017}.
Let $\Psi_{i,c}$ be the silhouette of $\mathcal{O}$, which is selected at each frame as in Sec.~\ref{sec:3d_pose_estimation}.
We compute the volume of $\mathcal{O}$ as Visual Hull (VH) \cite{Laurentini1994}, i.e.~the intersection of $C$ visual cones, each formed by projecting each $\Psi_{i,c}$ in 3D through the camera centre of $c$ \cite{Cheung2003}. 
Because finding visual cone intersections of generic 3D objects is not trivial, VH can be approximated by a set of voxels defining a 3D volume \cite{Mikhnevich2014}. 
Each voxel is projected on each camera image plane and if its 2D projection is within $\Psi_{i,c}$ for all $c \in \mathcal{C}$ then this voxel is considered to belong to $\mathcal{O}$'s VH.
In practice, due to noisy silhouette segmentation and/or noisy camera poses, one seeks if the voxel projection is within a subset of silhouettes rather than all.
Let $\mathcal{V}_i = \{\nu_{i,\theta} : \theta = 1, \dots, \Theta  \}$ be a set of voxels, where $\nu_{i,\theta} \in \mathbb{R}^3$ is the centre of the $\theta$th voxel approximating a volume of size $\Delta_\nu \times \Delta_\nu \times \Delta_\nu$ and $\Theta$ is the total number of voxels. 
We fix $\Delta_\nu$ to keep the same spatial resolution over time. 
Each $\nu_{i,\theta}$ is projected on each camera's image plane using the camera parameters $K_c$ and $P'_{i,c}$ for each $c$.
In our experiments we define a bounding volume composed of 160$\times$160$\times$160 voxels occupying a space of 1.8$\times$1.8$\times$1.9 meters centred in the position of the 3D skeleton's hip joint.
Let $x_{i,c,\theta}$ be the 2D projection of $\nu_{i,\theta}$ on $c$.
If $\lfloor x_{i,c,\theta} \rfloor \in \Psi_{i,c}$, where $\lfloor \cdot \rfloor$ is the rounding to lower integer operation, then $\nu_{i,\theta}$ is within the visual cone of $c$. 
If $\nu_{i,\theta}$ is within the visual cone of at least $n_\Psi$ cameras, then we deem $\nu_{i,\theta}$ to belong to $\mathcal{O}$'s VH.
We post-process VH with Marching Cubes to find the iso-surfaces and to produce the final mesh \cite{Lorensen1987_1}.

\section{Experimental results}\label{sec:results}

\subsection{Hardware and software}

The mobile app is developed in Unity3D \cite{unity3d}, using ARCore for SLAM \cite{arcore} and the Cloud Anchor to find $T_c$ \cite{cloudanchors}.
Capture sessions are managed by Unity3D Match Making \cite{matchmaking}.
The Relay Server and the Data Manager are embedded in separate Docker containers, running on a laptop within the local network.
In \cite{Bortolon2020} we describe our Relay Server architecture/protocols in detail.
JPEG frames are serialised/deserialised using Protobuf \cite{protobuf}.
We use the Sparse Bundle Adjustment C-library \cite{Lourakis2009}.
To collect data, we used two Huawei P20Pro, one OnePlus 5, one Samsung S9, one Xiaomi Pocophone F1 and one Xiaomi Mi8. 
Mobiles were connected to a 5GHz WiFi 802.11 access point.
We used a Raspberry Pi 2 to route packets to the internet.
Relay Server and Data Manager are deployed on an Edge computer within the same local network as the mobiles.

\subsection{Dataset}
We collected a dataset, namely 4DM, which involves six people recording a person acting table tennis (Fig.~\ref{fig:teaser}). 
We recorded the 4DM dataset outdoors, with cluttered backgrounds, cast shadows and with people appearing in each other's view, thus becoming likely distractors for object detection and human pose estimation.
We recorded three sequences: 4DM-Easy (312 frames, all recording people were static), 4DM-Medium (291 frames, three recording people were moving) and 4DM-Hard (329 frames, all recording people were moving and we added occluders to the scene).
The host generated triggers at 10Hz.
Frames have a resolution of 640$\times$480 and an average size of 160KB.
The latency between mobiles and the Relay Server was $\sim$5ms.

\subsection{Throughput and latency}

We compare our Edge solution with a Cloud-based one, i.e.~AWS \cite{aws}.
Tab.~\ref{tab:socket_comparison} shows the percentage of successfully received data in the case of Socket and HTTP based transmissions to the Cloud and to the Edge using 2.4GHz and 5GHz WiFi.
The host generated triggers at 20Hz for one minute.
HTTP is as effective as Socket when the bandwidth has enough capacity, but unreliable when there is limited capacity.
We observed that Socket-level data management is key to effectively manage large throughputs.
\begin{table}[t]
\centering
\caption{Percentage of successfully received data comparing architectures based on Socket and HTTP through Cloud and Edge services under different WiFi connections.}
\label{tab:socket_comparison}
\resizebox{.88\columnwidth}{!}{%
\begin{tabular}{lcccc}
\toprule
 \multirow{2}{*}{\textbf{Method}} & \multicolumn{2}{c}{\textbf{Cloud}} & \multicolumn{2}{c}{\textbf{Edge}} \\
  & 2.4GHz & 5GHz & 2.4GHz & 5GHz \\
 \midrule
 HTTP \cite{Bortolon2020} & 45.1\% & 78.2\% & 38.8\% & 79.0\% \\
 Ours & 59.2\% & 78.6\% & 49.3\% & 77.7\% \\
 \bottomrule
 \end{tabular}
}
\end{table}

\begin{figure}[t]
  \centering
  \includegraphics[width=1\columnwidth]{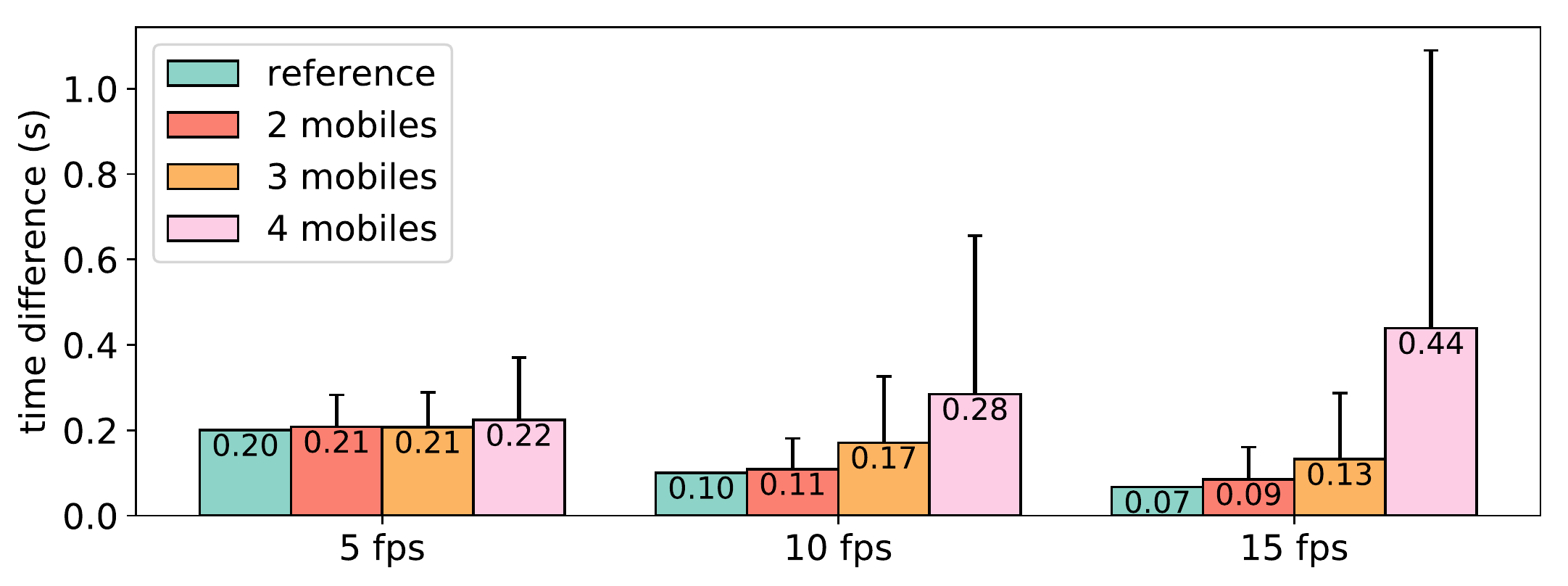}
  \caption{Average and standard deviation time difference between consecutive frames received on the Data Manager as a function of the trigger frequency and of the number of mobiles. Statistics are computed over three captures of 30s each. The reference is the ideal case.}
  \label{fig:frame_differences}
\end{figure}

Fig.~\ref{fig:frame_differences} shows the statistics of the time difference between consecutive frames received on the Data Manager as a function of the trigger frequency and the number of mobiles.
We used up to four mobiles connected to a 2.4GHz WiFi.
The host transmitted triggers at 5Hz, 10Hz and 15Hz for 30 seconds.
We associated a reference time difference for each capture showing the ideal case, e.g.~10Hz$\rightarrow$0.10s.
Experiments were repeated three times for each capture.
We can observe that when the network bandwidth is enough, the Data Manager can receive frames at the same trigger frequency as that set by the host mobile. When the throughput exceeds the bandwidth capacity, the system starts buffering packets due to TCP re-transmissions.

\subsection{End-to-end delay assessment}\label{sec:end-to-end_ass}

To assess the end-to-end delay we used two mobiles configured in the same way as in 4DM.
We adopted the same procedure as \cite{Bortolon2020} to quantify the frame capture delay between mobiles, i.e.~two mobiles captured the time ticked by a stopwatch (up to millisecond precision) displayed on a 60Hz computer screen (16ms refresh rate).
We extracted the time digits from each pair of frames captured by the two mobiles using OCR \cite{ocr} and computed their time difference.
Frames that were likely to provide incorrect OCR digits were manually corrected.
For each experiment, we used the first 100 captured frames and calculated the average and standard deviation of the difference between consecutive frames.
For comparison, we implemented the approach proposed in Sec.~IV of \cite{Ansari2019}, i.e.~the NTP is requested 50 times by each mobile and the average NTP over these requests is used as a global synchronisation between mobiles.
The clock of the mobiles is aligned to this estimated global time and used to synchronise the capture of each frame.
The host informs the other mobiles when to start the acquisition with respect to this estimated global time through a trigger.
As we explained in Sec.~\ref{sec:intro}, we cannot achieve a camera-sensor synchronisation as in \cite{Ansari2019}, because the capture of frames and the associated mobile poses depends on the SLAM process and the Unity3D rendering engine.
We name this approach NTP-Sec.~IV \cite{Ansari2019}.
We also implemented a NTP baseline version of NTP-Sec.~IV \cite{Ansari2019}, namely NTP-Baseline, where a single NTP request is used for the global time.

Tab.~\ref{tab:ocr_frame_difference} shows the comparison between our trigger-based synchronisation method and NTP-based synchronisation approaches.
We empirically measured a standard deviation of about 16ms in all cases, corresponding to the monitor refresh rate.
The results show that the synchronisation using our trigger-based method achieves the same accuracy as that of NTP-based approaches.
Because triggers are generated by the host, we can inherently vary their frequency online based on the application requirements.
For completeness, we also include the results reported in \cite{Ansari2019} in the case of mobiles synchronised without exploiting the prioritised camera sensor access.
The three ``Naive'' experiments used a single button press to trigger both phones to capture a frame each. 
The connectivity amongst mobiles was established through an audio wire, a Bluetooth connection and peer-to-peer WiFi.
Note that their measured delays are much higher than ours.

\begin{table}[t]
\caption{Statistics of the time differences measured between frames captured by two mobiles using different synchronisation methods. We reported the statistics measured by \cite{Ansari2019} and those we obtained using our OCR-based procedure.}
\label{tab:ocr_frame_difference}
\begin{center}
\resizebox{.9\columnwidth}{!}{%
    \begin{tabular}{llc}
    \toprule
    & \textbf{Method} & \textbf{Avg} $\boldsymbol{\pm}$ \textbf{Std} [ms]\\
    \midrule
    \multirow{3}{*}{\shortstack{Reported\\by~\cite{Ansari2019}}} & Naive Wired & $103\pm50$ \\
    & Naive Bluetooth & $69\pm65$ \\
    & Naive Wifi & $123\pm84$ \vspace{.05cm}\\
    \midrule
    \multirow{3}{*}{\shortstack{Measured\\by us\\with~OCR}} & NTP-Baseline & $29\pm16$\\
    & NTP-Sec.~IV \cite{Ansari2019} & $20\pm16$ \\
    & Ours & $21\pm17$\vspace{.05cm}\\
    \bottomrule
    \end{tabular}
   }
\end{center}
\end{table}

\subsection{3D skeleton reconstruction}\label{sec:reprojection_error}
\begin{figure}[b]
\begin{center}
  \begin{tabular}{@{}c@{}c}
    \begin{overpic}[width=.5\columnwidth]{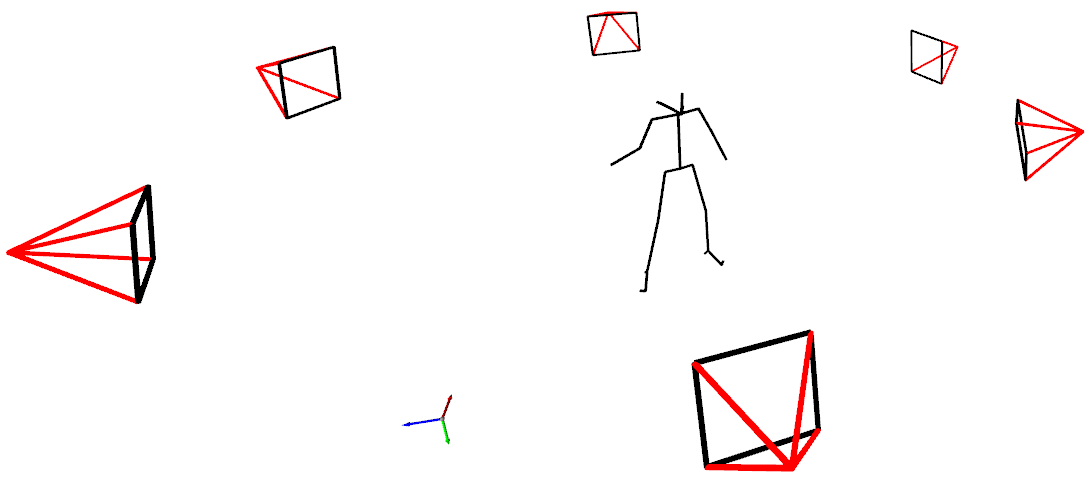}
      \put(0,0){\color{black}\scriptsize\textbf{frame 20}}
    \end{overpic}&
    \begin{overpic}[width=.5\columnwidth]{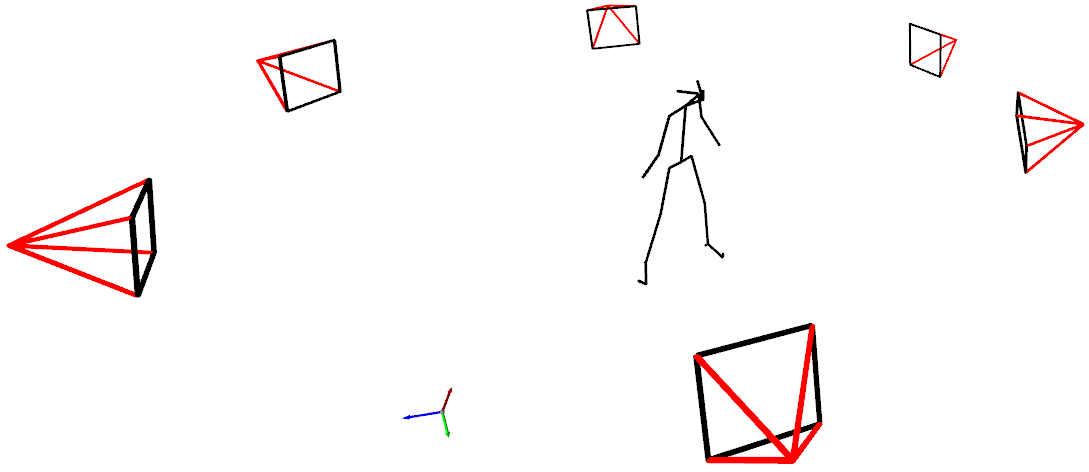}
      \put(0,0){\color{black}\scriptsize\textbf{frame 23}}
    \end{overpic}\\
    \begin{overpic}[width=.5\columnwidth]{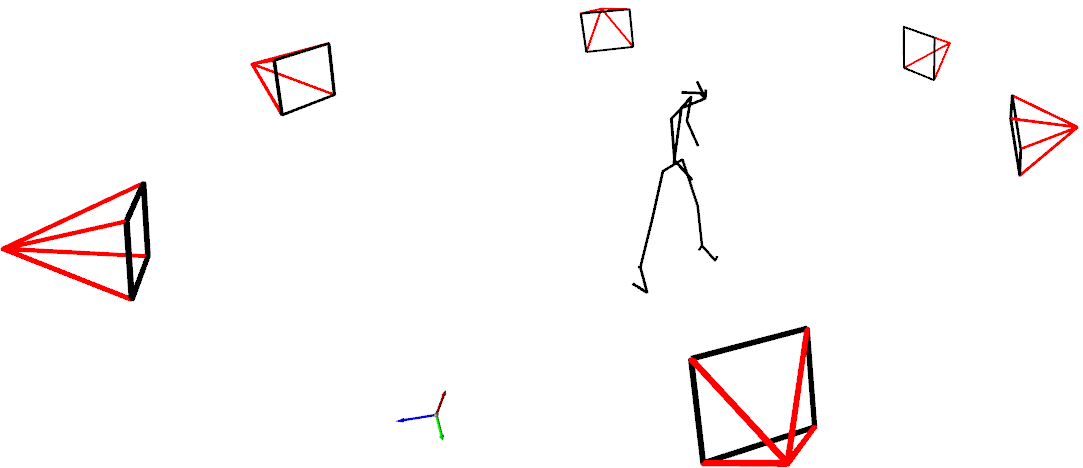}
      \put(0,0){\color{black}\scriptsize\textbf{frame 26}}
    \end{overpic}&
    \begin{overpic}[width=.5\columnwidth]{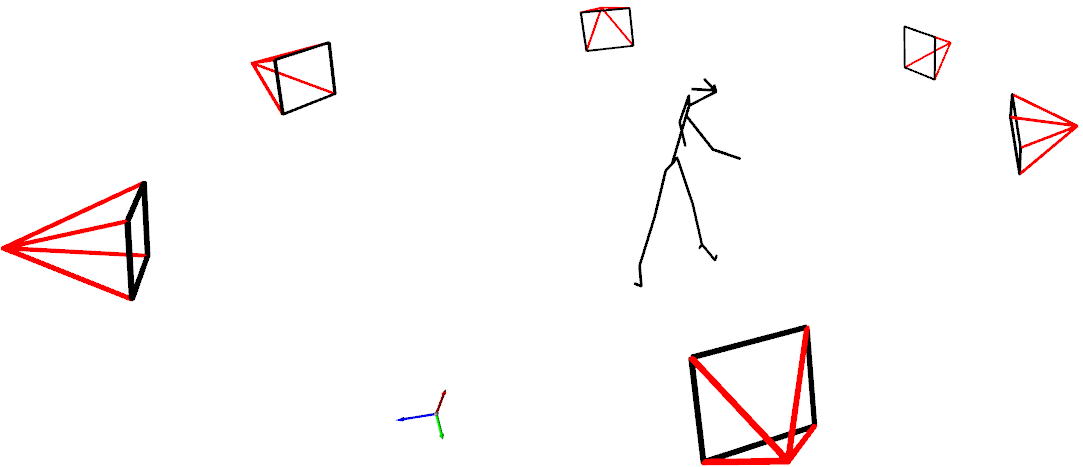}
      \put(0,0){\color{black}\scriptsize\textbf{frame 29}}
    \end{overpic}
  \end{tabular}
\end{center}
\caption{Sequence of frames of the reconstructed 3D skeleton from 4DM-Easy.}
\label{fig:sk_seq}
\end{figure}

\begin{figure}[t]
\begin{center}
  \begin{tabular}{@{}c@{}c@{}c}
    \includegraphics[width=1\columnwidth]{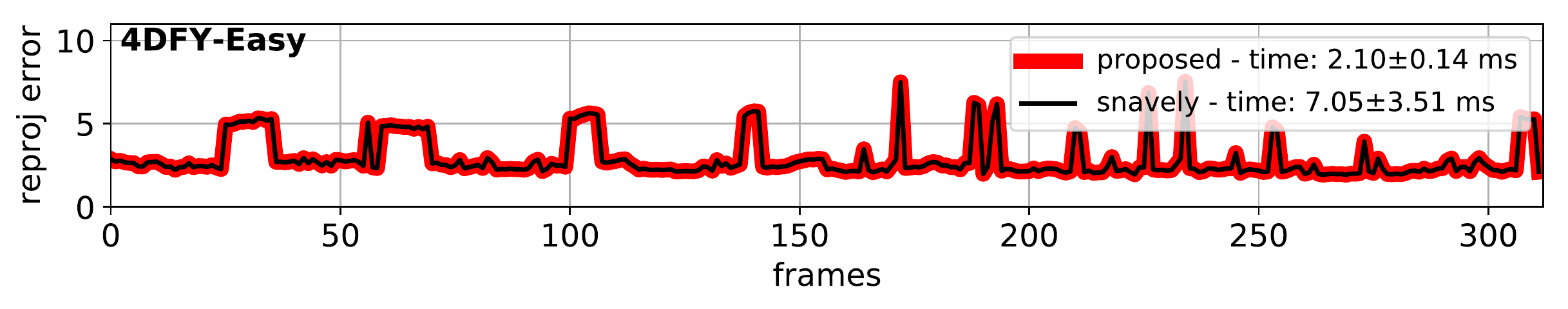}\\
    \includegraphics[width=1\columnwidth]{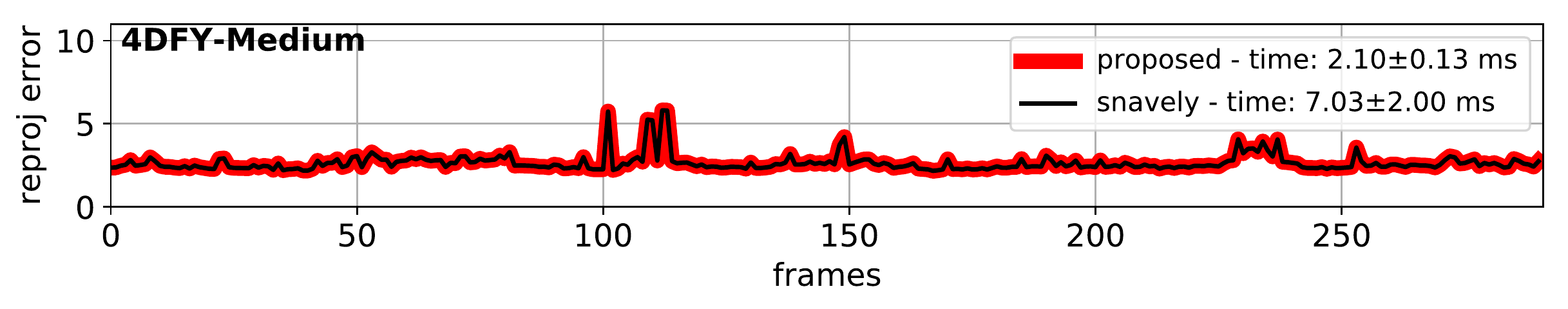}\\
    \includegraphics[width=1\columnwidth]{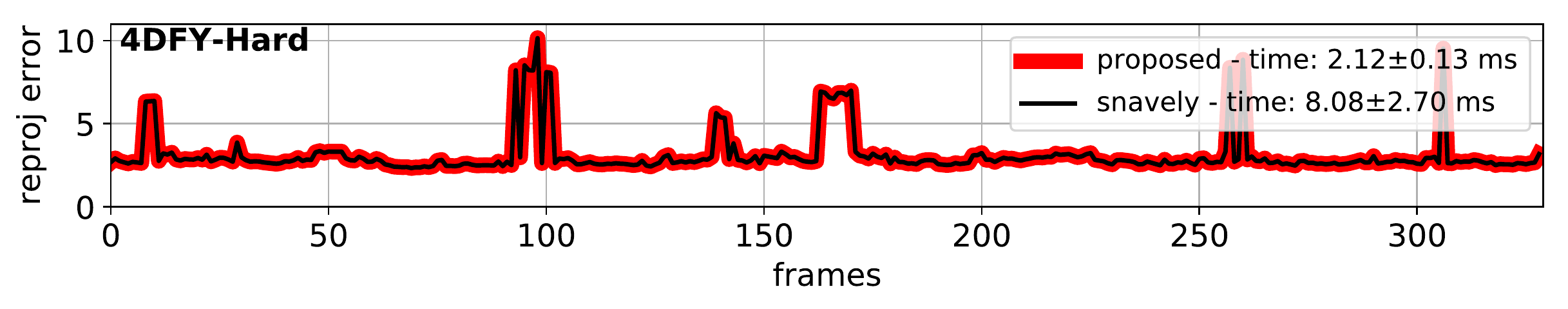}\\
  \end{tabular}
\end{center}
\caption{Comparison between Bundle Adjustment computed incrementally \cite{Snavely2007} and globally.
The re-projection error is reported in pixels over time for 4DM-Easy, 4DM-Medium and 4DM-Hard.}
\label{fig:resproj_err_4DM}
\end{figure}

\begin{figure}[t]
\begin{center}
  \begin{tabular}{@{}c}
    \includegraphics[width=1\columnwidth]{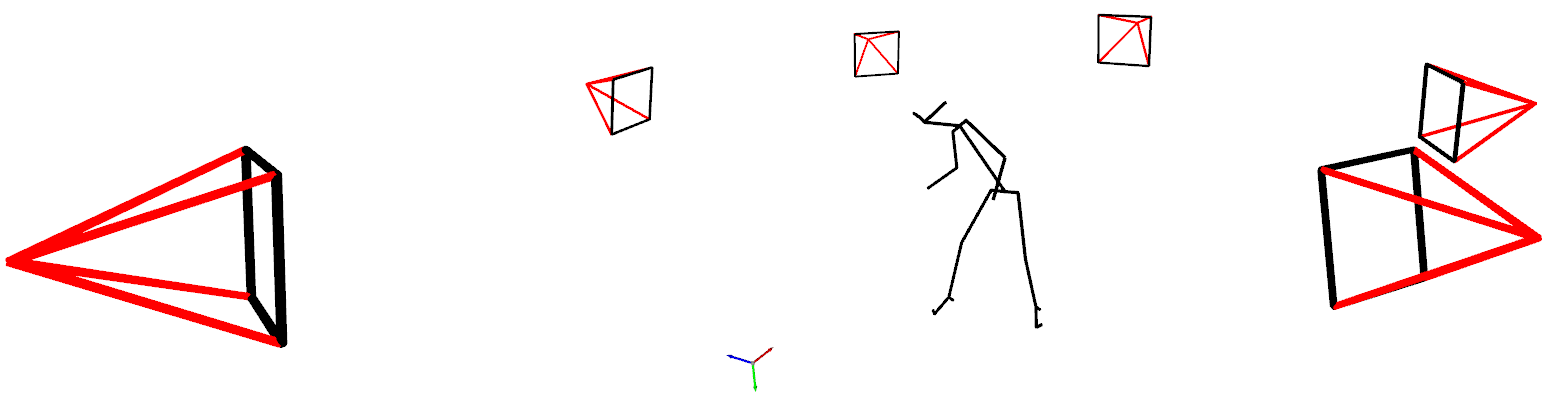}
  \end{tabular}
  \begin{tabular}{@{}c@{\,}c@{\,}c}
  \begin{overpic}[width=.33\columnwidth,height=.3\columnwidth]{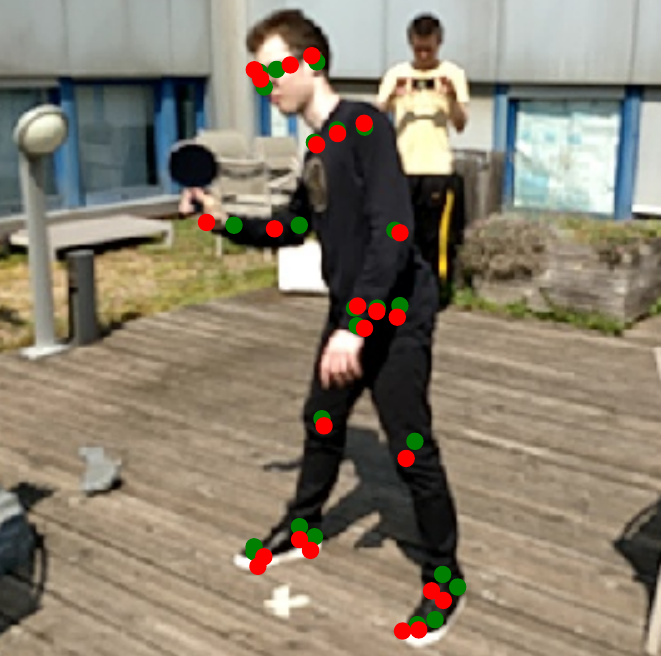}
      \put(2,3){\color{white}\footnotesize\textbf{mobile 1}}
    \end{overpic}
    \begin{overpic}[width=.33\columnwidth,height=.3\columnwidth]{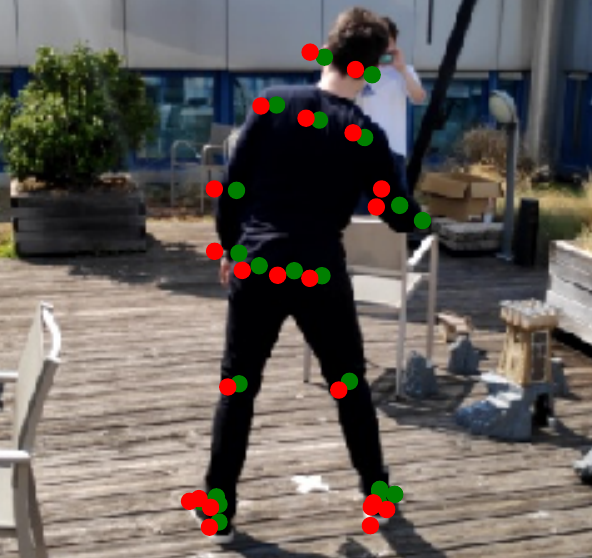}
      \put(2,3){\color{white}\footnotesize\textbf{mobile 3}}
    \end{overpic}
    \begin{overpic}[width=.33\columnwidth,height=.3\columnwidth]{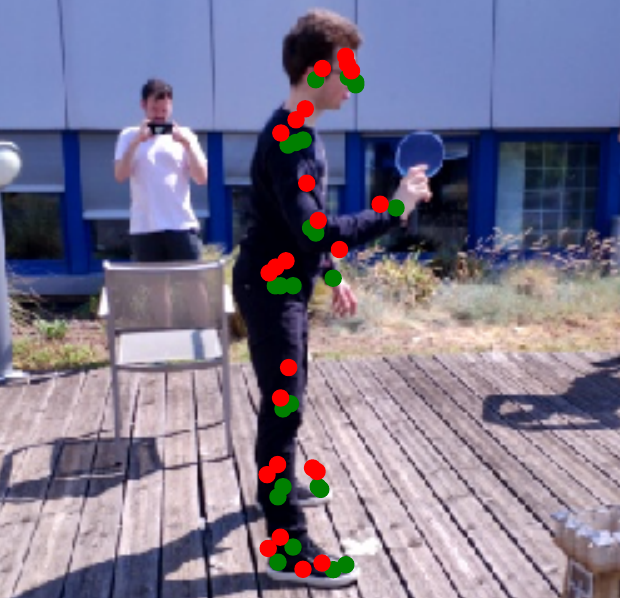}
      \put(2,3){\color{white}\footnotesize\textbf{mobile 5}}
    \end{overpic}
  \end{tabular}
\end{center}
\caption{3D skeleton and mobile poses in 4DM-Hard. Three zoomed-in frames overlaying the re-projected 3D joints (red dots) and the 2D joints computed with OpenPose (green dots).}
\label{fig:3d_pose_qualitative}
\end{figure}

Fig.~\ref{fig:sk_seq} shows a sequence of 3D skeletons reconstructed from 4DM-Easy, while Fig.~\ref{fig:resproj_err_4DM} reports the re-projection error using Eq.~\ref{eq:ba_initialisation}.
The average re-projection error (over time and across joints) is 2.86 pixels for 4DM-Easy, 2.64 pixels for 4DM-Medium and 3.11 pixels for 4DM-Hard.
We can observe that BA reaches the same solution as the strategy in \cite{Snavely2007}. 
However, as shown in the legends of Fig.~\ref{fig:resproj_err_4DM}, the global minimisation allows us to reduce by more than thrice the computational time for each $i$.
Fig.~\ref{fig:3d_pose_qualitative} shows the 3D skeleton and the mobile poses in 4DM-Hard at $i$=164, and three zoomed-in frames of the person captured from mobiles 1, 3 and 5, overlaying the re-projected 3D joints (red dots) and the 2D joints computed with OpenPose (green dots). 
Some of the re-projected 3D points are shifted from the 2D locations estimated with OpenPose.
We attribute these errors to the mobile poses that may have been inaccurately estimated by SLAM due to the high dynamicity of the scene (moving object and moving cameras), and to synchronisation delays.

\begin{figure}[t]
  \centering
  \includegraphics[width=1\columnwidth]{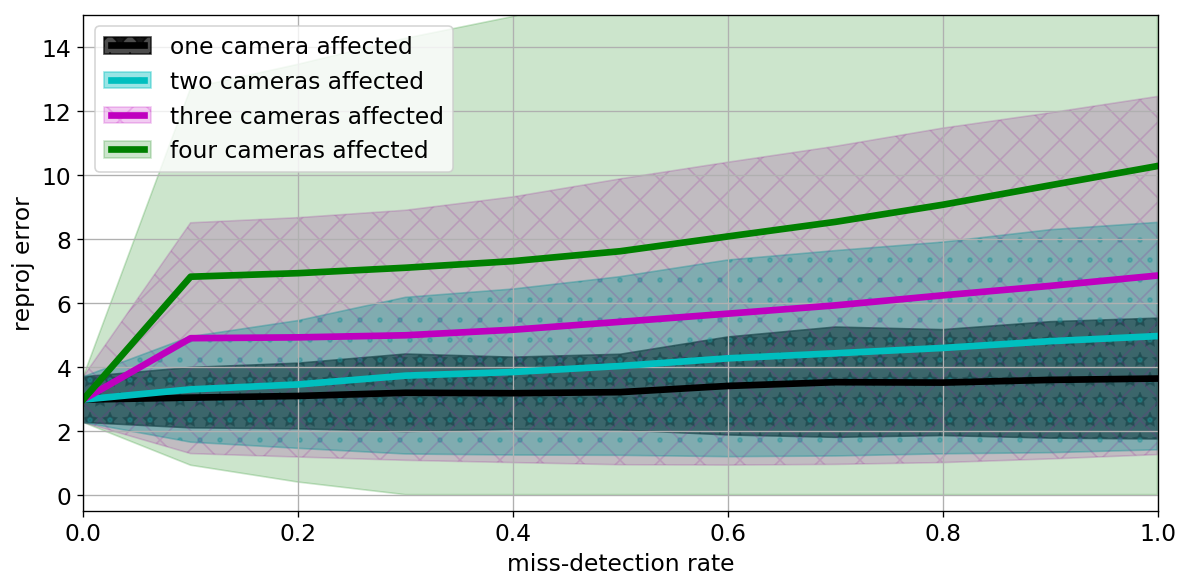}
  \caption{Sensitivity analysis showing the average (line) and the standard deviation (coloured area) of the re-projection error as a function of a simulated OpenPose's miss-detection rate affecting different numbers of cameras.}
  \label{fig:easy_missdets_reproj_err}
\end{figure}

To assess the robustness of our system, we simulated different miss-detection rates affecting OpenPose output on different numbers of cameras.
Miss-detection implies that the actor is not detected, hence the affected camera cannot contribute to triangulation.
Fig.~\ref{fig:easy_missdets_reproj_err} shows the average (line) and standard deviation (coloured area) of the re-projection error.
We can observe that miss-detections lead to an expected steady increase in re-projection error. 
We did not encounter breaking points at which the system completely fails triangulation.
As long as a valid camera pair (Sec.~\ref{sec:3d_pose_estimation}) does not undergo miss-detections, the system can triangulate the skeleton's joints.

\begin{figure*}[t]
\centering
\begin{tabular}{@{}c@{}c}
	\begin{tabular}{@{}c}
	\begin{overpic}[width=.87\columnwidth]{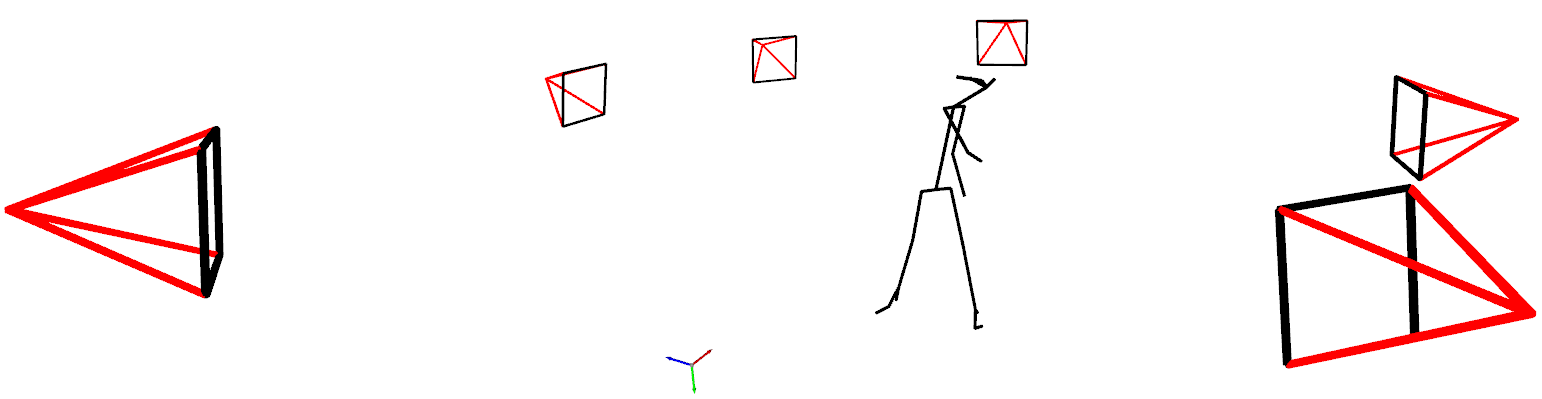}
	  \put(0,-3){\color{black}\scriptsize\textbf{bundle adjustment on 3D points}}
	\end{overpic}
	\vspace{.5cm}
	\\
	\begin{overpic}[width=.87\columnwidth]{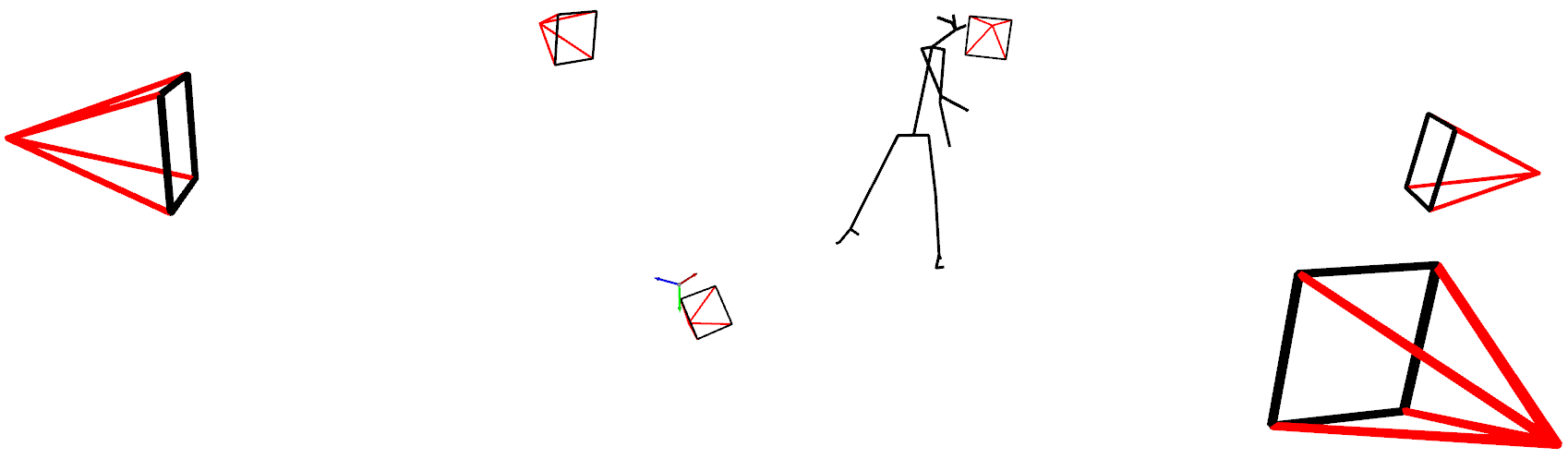}
	  \put(0,-5){\color{black}\scriptsize\textbf{bundle adjustment on 3D points and mobile poses}}
	\end{overpic}
	\end{tabular}
	\begin{tabular}{@{}c@{}c@{}c}
    \begin{overpic}[width=.37\columnwidth]{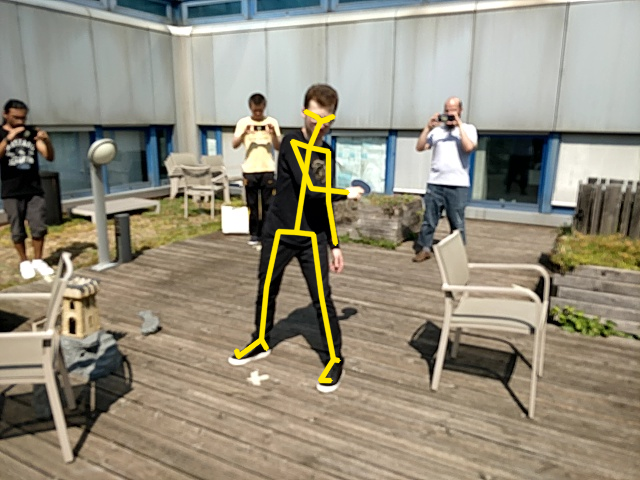}
      \put(2,3){\color{white}\footnotesize\textbf{mobile 1}}
    \end{overpic}&
    \begin{overpic}[width=.37\columnwidth]{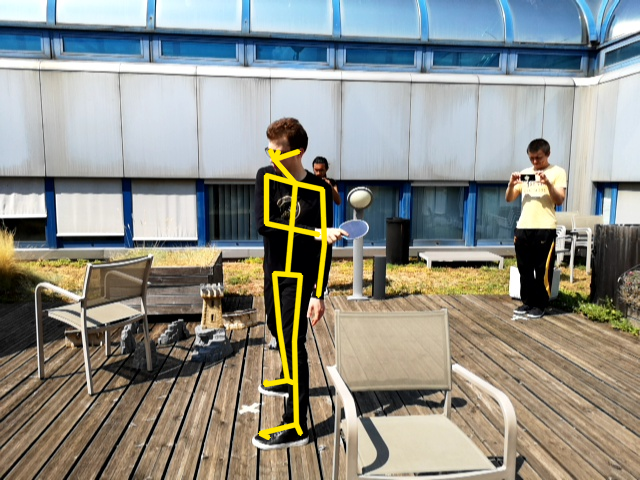}
      \put(2,3){\color{white}\footnotesize\textbf{mobile 2}}
    \end{overpic}&
    \begin{overpic}[width=.37\columnwidth]{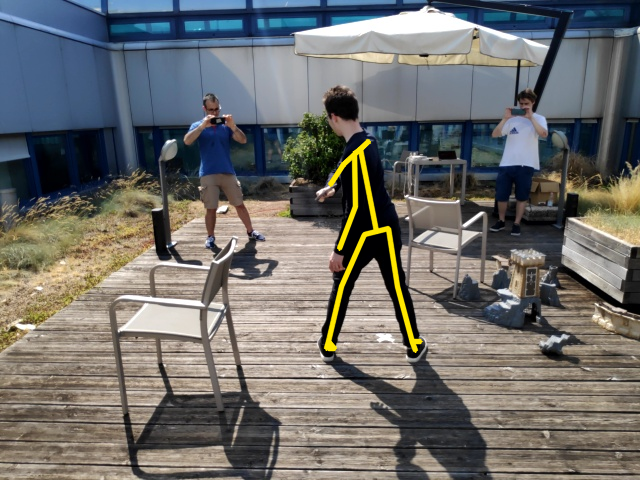}
      \put(2,3){\color{white}\footnotesize\textbf{mobile 3}}
    \end{overpic}\vspace{-.1cm}\\
    \begin{overpic}[width=.37\columnwidth]{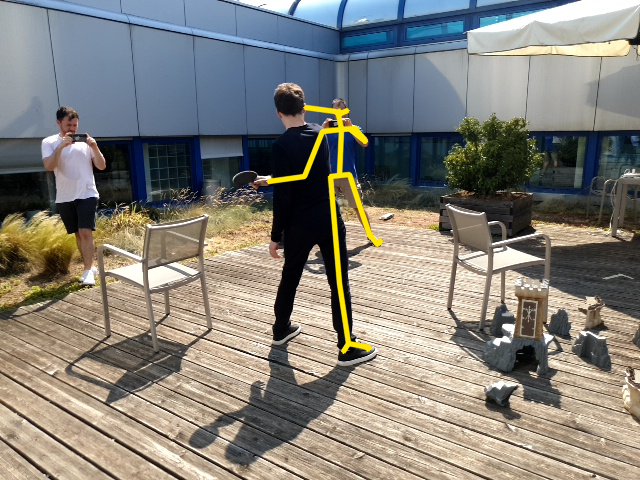}
      \put(2,3){\color{white}\footnotesize\textbf{mobile 4}}
      \put(0,-12){\color{black}\scriptsize\textbf{human poses estimated from each camera}}
    \end{overpic}&
    \begin{overpic}[width=.37\columnwidth]{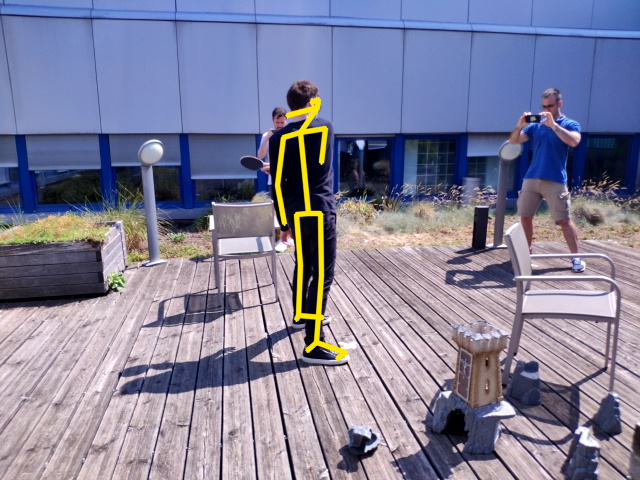}
      \put(2,3){\color{white}\footnotesize\textbf{mobile 5}}
    \end{overpic}&
    \begin{overpic}[width=.37\columnwidth]{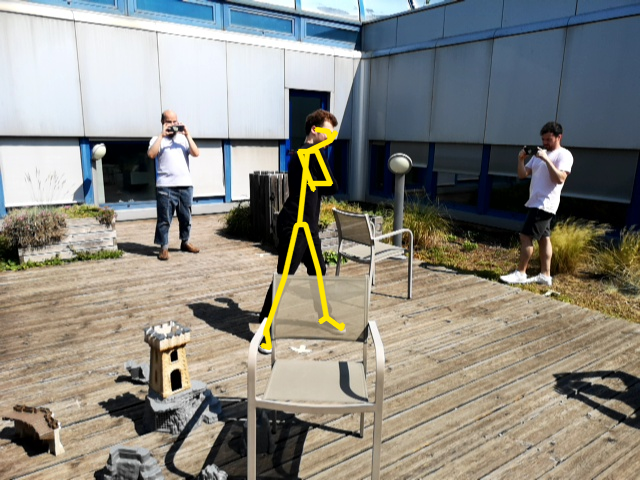}
      \put(2,3){\color{white}\footnotesize\textbf{mobile 6}}
    \end{overpic}\\
  \end{tabular}
\end{tabular}
\vspace{.4cm}
\caption{Reconstructed skeleton and mobile poses from 4DM-Hard. (Top-Left) Bundle Adjustment on 3D points only. (Bottom-Left) Bundle Adjustment on 3D points and mobile poses. (Right) 2D poses estimated on each captured frame: note OpenPose mistake in \emph{mobile 4} due to a distractor.}
\label{fig:comparison_ba}
\end{figure*}

BA can also jointly optimise camera parameters (i.e.~mobile poses) and 3D points.
Fig.~\ref{fig:comparison_ba} shows results (at $i$=143 in 4DM-Hard) by applying BA on 3D points only, and on both 3D points and mobile poses.
The average re-projection error in this example is 5.35 pixels for the former and 2.23 pixels for the latter.
The full sequence is available in the supplementary video.
Here we can observe that the 3D skeleton is estimated similarly in the two cases, but camera poses are different: they are often positioned in implausible locations.
To further investigate this issue, we fixed one mobile as world reference and optimised the others in order to enforce BA to anchor the optimisation with respect to this mobile's pose (reference), without seeking a new coordinate re-arrangement.
However, we still experienced this undesired behaviour.
We believe that this happens for three reasons.
First, there are similar numbers of 3D points and camera parameters to optimise.
This may hinder BA in converging to the correct solution, as the structure of the environment has few 3D points.
Second, mobiles may receive triggers with different delays. 
Delays may vary based on the quality of the wireless connection: e.g.~if an external user, within the same wireless network, generates traffic (e.g.~downloads a file), they will affect the network latency.
Third, BA assumes 2D joints to be correct. 
Hence, if OpenPose fails to estimate some joints, this will then affect the minimisation process.
An example is in \emph{mobile 4} of Fig.~\ref{fig:comparison_ba}, where the person behind the actor has affected OpenPose estimation.

\subsection{Comparison with 3D pose estimation from 2D}\label{sec:comparison2Dto3D}
We compare the quality of our reconstructed 3D skeleton with the 3D skeleton estimated with a deep-learning-based method, namely VideoPose3D \cite{Pavllo2019}, that infers 3D skeletons directly from 2D images.
VideoPose3D processes the 2D joints computed on the images of the whole sequence and outputs the respective 3D joints for each frame with respect to the mobile coordinate system.
VideoPose3D exploits the temporal information to generate 3D joint locations over time.
The input 2D joints are computed with Detectron2 \cite{wu2019detectron2}.
Differently from OpenPose, the skeleton model of VideoPose3D uses $M$=15 joints.
We use a model pretrained on Human3.6M \cite{Ionescu2014}.
To make the results comparable, we input the same 2D joints estimated with Detectron2 to our system.

\begin{figure}[t]
\begin{center}
  \begin{tabular}{@{}c@{\,\,\,\,\,\,}c}
  	\begin{overpic}[width=.4\columnwidth]{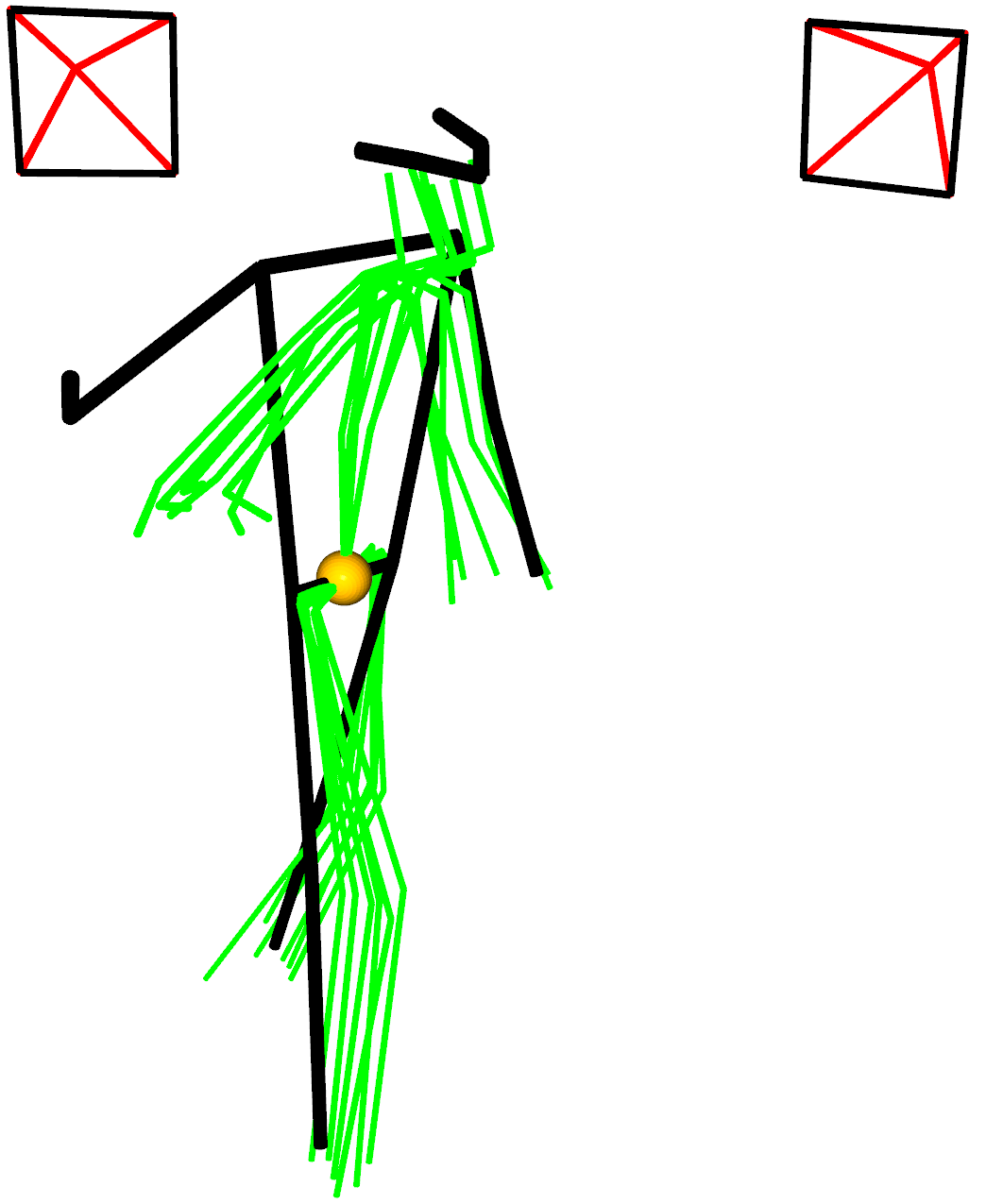}
      \put(0,0){\color{black}\scriptsize\textbf{(a)}}
    \end{overpic}&
    \begin{overpic}[width=.4\columnwidth]{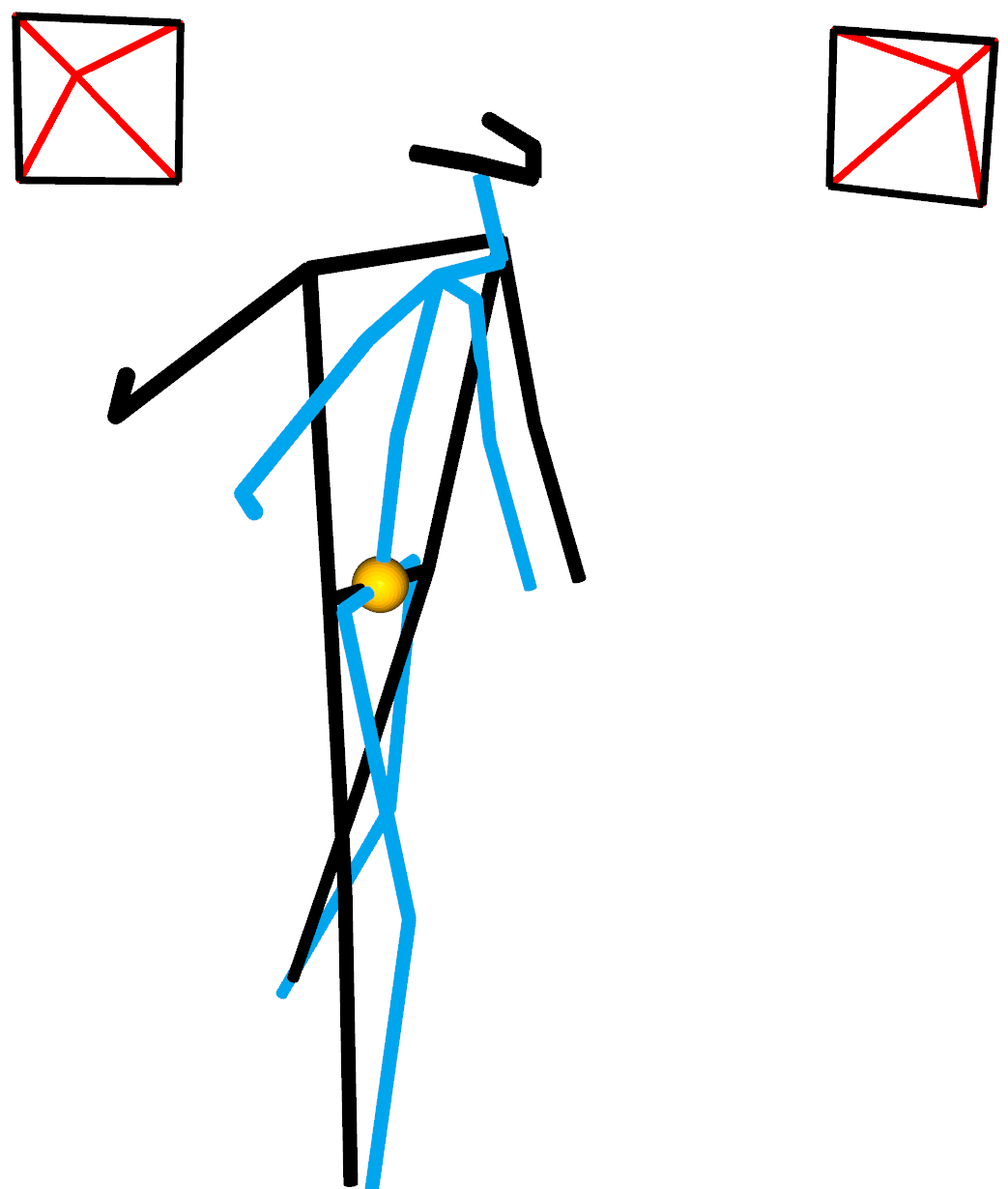}
      \put(0,1){\color{black}\scriptsize\textbf{(b)}}
    \end{overpic}
  \end{tabular}
\end{center}
\caption{Overlay between the 3D skeleton triangulated with our method (black) and the skeletons obtained with VideoPose3D \cite{Pavllo2019} opportunely rescaled (green and light blue). (a) Six 3D skeletons estimated from each mobile with VideoPose3D (green), and (b) the skeleton whose joints are computed as the centres of mass amongst the corresponding joints inferred from all the mobiles with VideoPose3D (light blue). The yellow sphere indicates the point used to make the VideoPose3D's output skeleton coincide with that estimated with our system.}
\label{fig:videopose_sk_comparison}
\end{figure}
Fig.~\ref{fig:videopose_sk_comparison} shows the overlay between the 3D skeleton triangulated with our method (black), (a) the six 3D skeletons estimated from each mobile with VideoPose3D (green), and (b) the skeleton whose joints are computed as the centres of mass amongst the corresponding joints inferred from all the mobiles with VideoPose3D (light blue).
VideoPose3D's output skeleton is up to a scale.
To overlay the skeleton obtained via triangulation with our system with that of VideoPose3D, we scaled VideoPose3D's skeleton by making the hip centres of the two skeletons coincide (see the yellow spheres).
VideoPose3D outputs a different torso and head's representations of the input skeleton.
Therefore we compute the re-projection error only for the skeletons' joints that exist in both input and output.
VideoPose3D is an offline approach as it uses temporal information: the skeleton pose in a frame is computed using past and future information.
Although the skeletons that are estimated with the two methods appear with a similar pose, the joints do not coincide.
In this regard Tab.~\ref{tab:reproj_err_videopose} reports the re-projection error.
The re-projection error of VideoPose3D is consistently higher than that of our system in all three setups.

\begin{table}[t]
\caption{Comparison between the 3D skeleton reconstructed with our system and that estimated with VideoPose3D \cite{Pavllo2019} in terms of re-projection error (average and standard deviation calculated over all the frames of each dataset).}
\label{tab:reproj_err_videopose}
\begin{center}
\resizebox{.8\columnwidth}{!}{%
    \begin{tabular}{llc}
    \toprule
    & \textbf{Method} & \textbf{Avg} $\boldsymbol{\pm}$ \textbf{Std} [pixels]\\
    \midrule
    \multirow{2}{*}{\rotatebox[origin=c]{90}{\parbox[c]{.5cm}{\centering Easy}}} & VideoPose3D \cite{Pavllo2019} & $6.836\pm1.303$ \\
    & Ours & $4.735\pm1.195$ \\
    \midrule
    \multirow{2}{*}{\rotatebox[origin=c]{90}{\parbox[c]{.5cm}{\centering Med}}} & VideoPose3D \cite{Pavllo2019} & $7.403\pm1.130$ \\
    & Ours & $5.053\pm1.072$ \\
    \midrule
    \multirow{2}{*}{\rotatebox[origin=c]{90}{\parbox[c]{.5cm}{\centering Hard}}} & VideoPose3D \cite{Pavllo2019} & $7.803\pm1.218$ \\
    & Ours & $5.645\pm1.316$ \\
    \bottomrule
    \end{tabular}
   }
\end{center}
\end{table}

\subsection{Volumetric reconstruction}\label{sec:res_volum_recon}
\begin{figure}[t]
\begin{center}
  \begin{tabular}{@{}c}
  	\begin{overpic}[width=1\columnwidth]{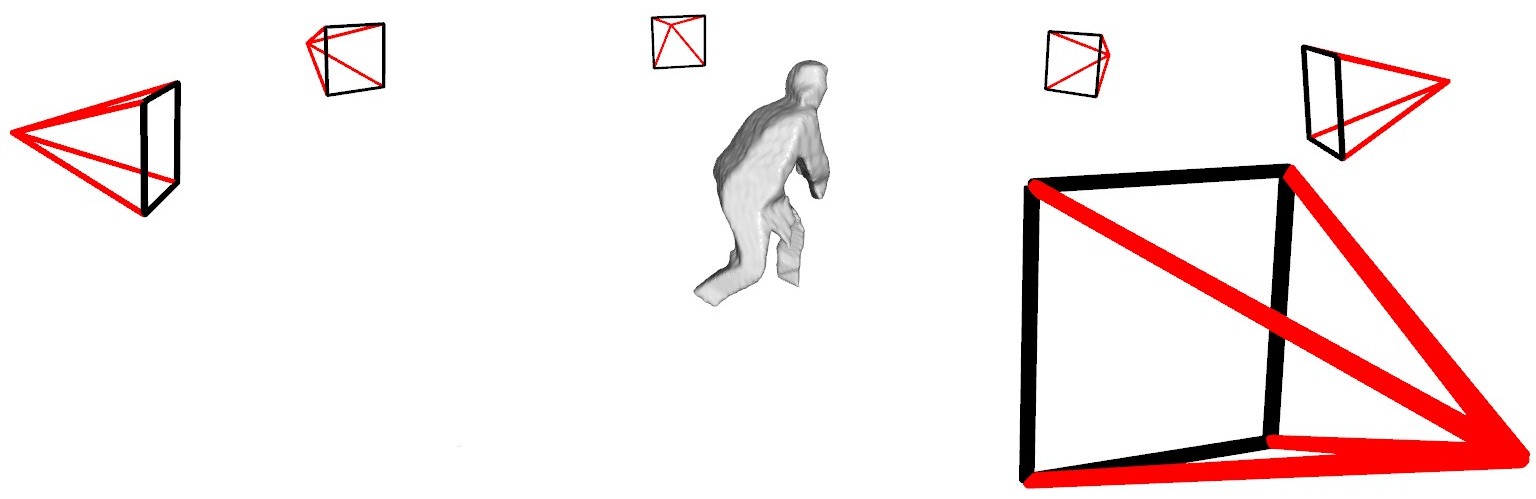}
      \put(0,0){\color{black}\scriptsize\textbf{4DM-Easy}}
    \end{overpic}\\
    \begin{overpic}[width=1\columnwidth]{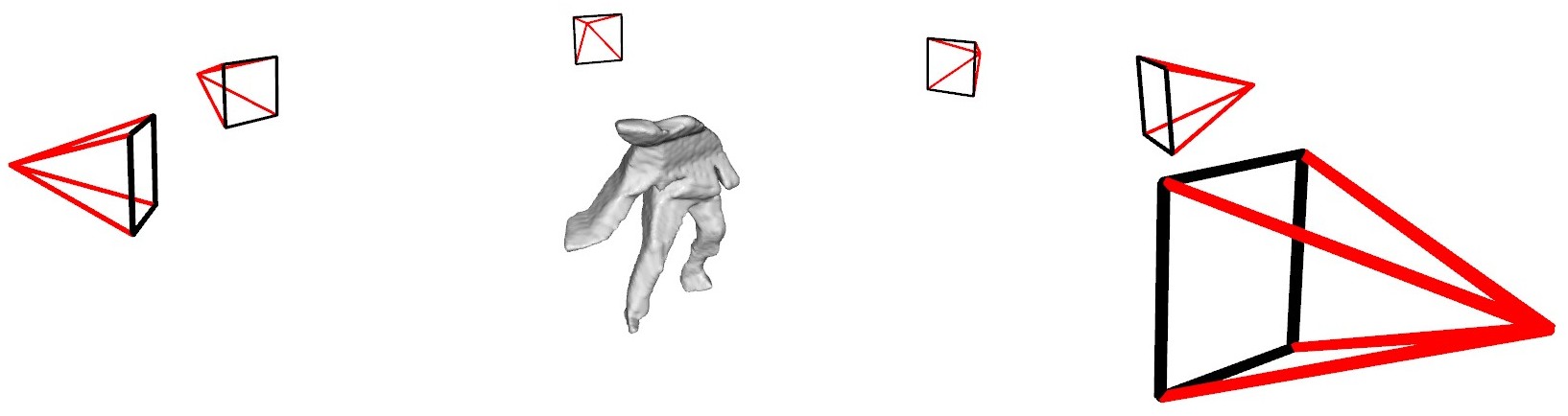}
      \put(0,0){\color{black}\scriptsize\textbf{4DM-Medium}}
    \end{overpic}\\
    \begin{overpic}[width=1\columnwidth]{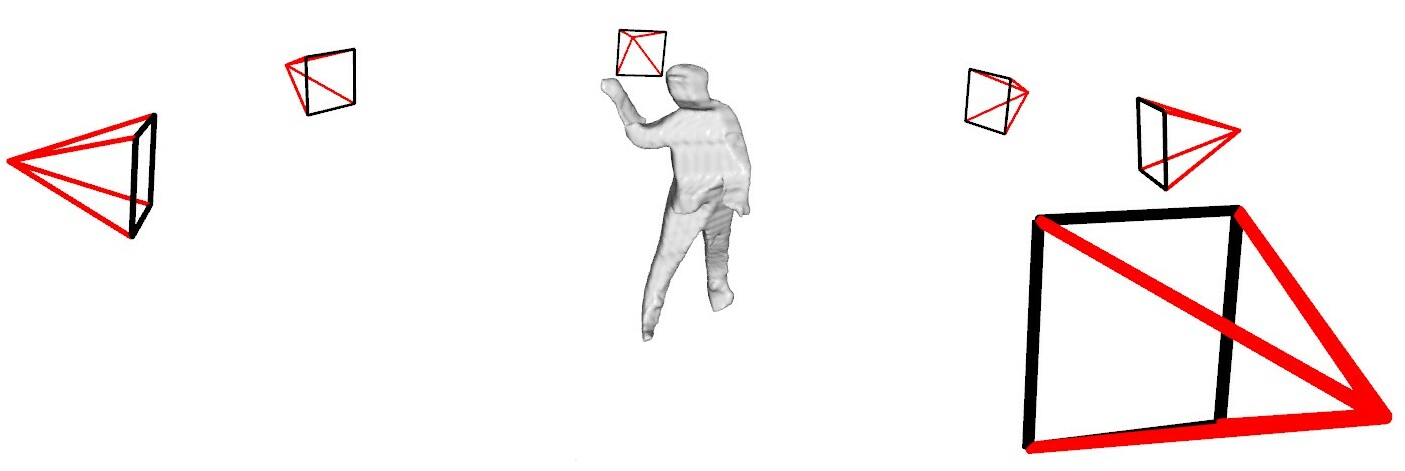}
      \put(0,0){\color{black}\scriptsize\textbf{4DM-Hard}}
    \end{overpic}
  \end{tabular}
\end{center}
\caption{Examples of volumetric reconstruction obtained with a shape-from-silhouette-based approach applied to our dataset 4DM. Silhouettes are automatically extracted using PointRend \cite{Kirillov2020}. Although silhouettes are not accurately following the outline of the actor, these results show that mobile poses and frames are captured synchronously. Shape-from-silhouette approaches are typically sensitive to noisy camera poses and synchronisation.}
\label{fig:vol_recon_1}
\end{figure}

Fig.~\ref{fig:vol_recon_1} shows examples of volumetric reconstructions extracted from 4DM-Easy, 4DM-Medium and 4DM-Hard.
The supplementary video contains the full sequences.
Although we are using a fully automated volumetric reconstruction baseline, we can observe that we can achieve promising reconstruction results even when cameras are moving.
Note that the quality of the reconstructed mesh is limited by the number of cameras: in fact, we can distinguish the intersection of the projected cones.
This is a known problem of shape-from-silhouette based approaches.
To obtain reconstructions with higher geometric details there should be a larger number of cameras or more sophisticated approaches, as those proposed in \cite{Pages2018,Mustafa2019}, should be used.

\section{Conclusions}\label{sec:concl}

We presented the first mobile-based system designed for real-time synchronous image captures that is based on commodity hardware.
We exploit a Relay Server to handle synchronisation triggers and a Data Manager to retrieve data (frames and mobile poses) efficiently.
These two modules can be deployed at the Edge or in the Cloud as it uses conventional wireless connections.
We showed that our system can effectively capture nearly-synchronous frames that can be used to reconstruct the 3D skeleton and the mesh of a moving person over time.
We empirically observed that mobile poses are sometimes inaccurate and that data may undergo variable delays.
A limitation of our system is that non-expert users may capture a dynamic object from ineffective viewpoints, for example ending up with small baselines due to mobiles moving close to each other.
An improvement could be the implementation of a AR-based UI/UX to guide non-experts during data captures.
Our delay compensation strategy currently computes the average RTT over all the past measurements and does not yet take into consideration RTT variations that may occur during a capture session. A future improvement of this can be the extension of the delay compensation strategy by continuously monitoring the RTT between host and clients, and updating the average RTT during capture using past measurements within a temporal window.
Other future improvements include on-line camera recalibration, on-line latency estimation to achieve a more robust 3D reconstruction, and validation on a 5G network infrastructure.

\begin{acknowledgements}
This research has received funding from the Fondazione CARITRO - Ricerca e Sviluppo programme 2018-2020.
\end{acknowledgements}

\bibliographystyle{spmpsci}
\bibliography{refs}

\begin{thebibliography}{10}
\providecommand{\url}[1]{{#1}}
\providecommand{\urlprefix}{URL }
\expandafter\ifx\csname urlstyle\endcsname\relax
  \providecommand{\doi}[1]{DOI~\discretionary{}{}{}#1}\else
  \providecommand{\doi}{DOI~\discretionary{}{}{}\begingroup
  \urlstyle{rm}\Url}\fi

\bibitem{arcore}
{ARCore}.
\newblock \url{https://developers.google.com/ar} (Accessed: Mar 2020)

\bibitem{ocr}
{Amazon Textract}.
\newblock \url{https://aws.amazon.com/textract/} (Accessed: Oct 2020)

\bibitem{aws}
{Amazon Web Services}.
\newblock \url{https://aws.amazon.com} (Accessed: Oct 2020)

\bibitem{cloudanchors}
{Cloud Anchors}.
\newblock
  \url{https://developers.google.com/ar/develop/java/cloud-anchors/overview-android#how-hosted}
  (Accessed: Oct 2020)

\bibitem{mlapi_relay}
{MLAPI Relay}.
\newblock \url{https://github.com/MidLevel/MLAPI.Relay} (Accessed: Oct 2020)

\bibitem{protobuf}
{Protocol Buffers}.
\newblock \url{https://developers.google.com/protocol-buffers/} (Accessed: Oct
  2020)

\bibitem{unity3d}
{Unity3D}.
\newblock \url{https://unity.com/} (Accessed: Oct 2020)

\bibitem{matchmaking}
{Unity3D MatchMaking}.
\newblock
  \url{https://docs.unity3d.com/520/Documentation/Manual/UNetMatchMaker.html}
  (Accessed: Oct 2020)

\bibitem{Ansari2019}
Ansari, S., Wadhwa, N., R, G., Chen, J.: {Wireless Software Synchronization of
  Multiple Distributed Cameras}.
\newblock In: Proc. of IEEE ICCP. Tokyo, JP (2019)

\bibitem{Ballan2010}
Ballan, L., Brostow, G., Puwein, J., Pollefeys, M.: Unstructured video-based
  rendering: Interactive exploration of casually captured videos.
\newblock In: Proc. of SIGGRAPH. Los Angeles, US (2010)

\bibitem{Bortolon2020}
Bortolon, M., Chippendale, P., Messelodi, S., Poiesi, F.: {Multi-view data
  capture using edge-synchronised mobiles}.
\newblock In: Proc. of Computer Vision, Imaging and Computer Graphics Theory
  and Applications. Valletta, MT (2020)

\bibitem{Cao2017}
Cao, Z., Simon, T., Wei, S.E., Sheikh, Y.: {Realtime Multi-Person 2D Pose
  Estimation using Part Affinity Fields}.
\newblock In: Proc. of IEEE CVPR. Honolulu, US (2017)

\bibitem{Cheung2003}
Cheung, G.: Visual hull construction, alignment and refinement for human
  kinematic modeling, motion tracking and rendering.
\newblock {PhD} thesis, Carnegie Mellon University (2003)

\bibitem{Collet2015}
Collet, A., Chuang, M., Sweeney, P., Gillett, D., Evseev, D., Calabrese, D.,
  Hoppe, H., Kirk, A., Sullivan, S.: High-quality streamable free-viewpoint
  video.
\newblock ACM Trans. on Graphics \textbf{34}(4), 1--13 (2015)

\bibitem{Garg2018}
Garg, A., Yadav, A., Sikora, A., Sairam, A.: {Wireless Precision Time
  Protocol}.
\newblock IEEE Communication Letters \textbf{22}(4), 812--815 (2018)

\bibitem{Guillemaut2011}
Guillemaut, J.Y., Hilton, A.: {Joint Multi-Layer Segmentation and
  Reconstruction for Free-Viewpoint Video Applications}.
\newblock International Journal on Computer Vision \textbf{93}(1), 73--100
  (2011)

\bibitem{Hartley}
Hartley, R., Zisserman, A.: {Multiple View Geometry in Computer Vision}.
\newblock Cambridge University Press (2003)

\bibitem{He2017}
He, K., Gkioxari, G., Dollar, P., Girshick, R.: {Mask R-CNN}.
\newblock In: Proc. of IEEE ICCV. Venice, IT (2017)

\bibitem{Hu2016}
Hu, Y., Niu, D., Li, Z.: {A Geometric Approach to Server Selection for
  Interactive Video Streaming}.
\newblock IEEE Trans. on Multimedia \textbf{18}(5), 840--851 (2016)

\bibitem{Huang2014}
Huang, C.H., Boyer, E., Navab, N., Ilic, S.: Human shape and pose tracking
  using keyframes.
\newblock In: Proc. of IEEE CVPR. Columbus, US (2014)

\bibitem{Ionescu2014}
Ionescu, C., Papava, D., Olaru, V., Sminchisescu, C.: {Human3.6M: Large Scale
  Datasets and Predictive Methods for 3D Human Sensing in Natural
  Environments}.
\newblock IEEE Trans. on PAMI \textbf{36}(7) (2014)

\bibitem{Kim2012}
Kim, H., Guillemaut, J.Y., Takai, T., Sarim, M., Hilton, A.: {Outdoor Dynamic
  3D Scene Reconstruction}.
\newblock IEEE Trans. on Circuits and Systems for Video Technology
  \textbf{22}(11), 1611--1622 (2012)

\bibitem{Kirillov2020}
Kirillov, A., Wu, Y., He, K., Girshick, R.: {PointRend: Image Segmentation as
  Rendering}.
\newblock In: Proc. of IEEE CVPR. Seattle, US (2020)

\bibitem{Latimer2015}
Latimer, R., Holloway, J., Veeraraghavan, A., Sabharwal, A.: {SocialSync:
  Sub-Fr\cite{Ansari2019}e Synchronization in a Smartphone Camera Network}.
\newblock In: Proc. of ECCV Workshops. Zurich, CH (2015)

\bibitem{Laurentini1994}
Laurentini, A.: The visual hull concept for silhouette-based image
  understanding.
\newblock IEEE Trans. on PAMI \textbf{16} (1994)

\bibitem{Leo2008}
Leo, M., Mosca, N., Spagnolo, P., Mazzeo, P., D'Orazio, T., Distante, .A.:
  Real-time multiview analysis of soccer matches for understanding interactions
  between ball and players.
\newblock In: Proc. of Content-based Image and Video Retrieval. Niagara Falls,
  CA (2008)

\bibitem{Lorensen1987_1}
Lorensen, W., Cline, H.: {Marching Cubes: A High Resolution 3D Surface
  Construction Algorithm}.
\newblock In: Proc. of ACM SIGGRAPH. Anaheim, US (1987)

\bibitem{Lourakis2009}
Lourakis, M.A., Argyros, A.: {SBA: A Software Package for Generic Sparse Bundle
  Adjustment}.
\newblock ACM Trans. Math. Software \textbf{36}(1), 1--30 (2009)

\bibitem{Mikhnevich2014}
Mikhnevich, M., Hebert, P., Laurendeau, D.: Unsupervised visual hull extraction
  in space, time and light domains.
\newblock Computer Vision and Image Understanding \textbf{125}, 55--71 (2014)

\bibitem{Mills1991}
Mills, D.: {Internet time synchronization: the network time protocol}.
\newblock IEEE Trans. on Communications \textbf{39}(10) (1991)

\bibitem{Mustafa2017}
Mustafa, A., Hilton, A.: Semantically coherent co-segmentation and
  reconstruction of dynamic scenes.
\newblock In: Proc. of IEEE CVPR. Honolulu, US (2017)

\bibitem{Mustafa2019}
Mustafa, A., Russell, C., Hilton, A.: {U4D: Unsupervised 4D Dynamic Scene
  Understanding}.
\newblock In: Proc. of IEEE ICCV. Seoul, KR (2019)

\bibitem{Mustafa2019b}
Mustafa, A., Volino, M., Kim, H., Guillemaut, J.Y., Hilton, A.: Temporally
  coherent general dynamic scene reconstruction.
\newblock International Journal in Computer Vision  (2019)

\bibitem{Pages2018}
Pages, R., Amplianitis, K., Monaghan, D., andA. Smolic, J.O.: {Affordable
  content creation for free-viewpoint video and VR/AR applications}.
\newblock Journal of Vis. Comm. and Im. Repr. \textbf{53} (2018)

\bibitem{Pavllo2019}
Pavllo, D., Feichtenhofer, C., Grangier, D., Auli, M.: {3D human pose
  estimation in video with temporal convolutions and semi-supervised training}.
\newblock In: Proc. of IEEE CVPR. Long Beach, US (2019)

\bibitem{Poiesi2017}
Poiesi, F., Locher, A., Chippendale, P., Nocerino, E., Remondino, F., Gool,
  L.V.: {Cloud-based Collaborative 3D Reconstruction Using Smartphones}.
\newblock In: Proc. of European Conference on Visual Media Production (2017)

\bibitem{Pretto2017}
Pretto, N., Poiesi, F.: Towards gesture-based multi-user interactions in
  collaborative virtual environments.
\newblock In: Int. Arch. Photogramm. Remote Sens. Spatial Inf. Sci., XLII-2/W8.
  Hamburg, GE (2017)

\bibitem{Qiao2019}
Qiao, X., Ren, P., Dustdar, S., Liu, L., Ma, H., Chen, J.: {Web AR: A Promising
  Future for Mobile Augmented Reality--State of the Art, Challenges, and
  Insights}.
\newblock Proc. of the IEEE \textbf{107}(4), 651--666 (2019)

\bibitem{Rematas2018}
Rematas, K., Kemelmacher-Shlizerman, I., Curless, B., Seitz, S.: Soccer on your
  tabletop.
\newblock In: Proc. of IEEE Computer Vision and Pattern Recognition. Salt Lake
  City, US (2018)

\bibitem{Richardt2016}
Richardt, C., Kim, H., Valgaerts, L., Theobalt, C.: Dense wide-baseline scene
  flow from two handheld video cameras.
\newblock In: Proc. of 3DV. San Francisco, US (2016)

\bibitem{Sigal2009}
Sigal, L., Balan, A., Black, M.: Humaneva: Synchronized video and motion
  capture dataset and baseline algorithm for evaluation of articulated human
  motion.
\newblock International Journal of Computer Vision \textbf{87}(4) (2009)

\bibitem{Slabaugh2001}
Slabaugh, G., Schafer, R., Malzbender, T., Culbertson, B.: A survey of methods
  for volumetric scene reconstruction from photographs.
\newblock In: Proc. of Eurographics. Manchester, UK (2001)

\bibitem{Snavely2007}
Snavely, N., Seitz, S., Szeliski, R.: {Modeling the World from Internet Photo
  Collections}.
\newblock International Journal of Computer Vision \textbf{80}(2), 189--210
  (2007)

\bibitem{Soret2014}
Soret, B., Mogensen, P., Pedersen, K., Aguayo-Torres, M.: {Fundamental
  tradeoffs among reliability, latency and throughput in cellular networks}.
\newblock In: IEEE Globecom Workshops. Austin, US (2014)

\bibitem{Vo2016}
Vo, M., Narasimhan, S., Sheikh, Y.: {Spatiotemporal Bundle Adjustment for
  Dynamic 3D Reconstruction}.
\newblock In: Proc. of IEEE CVPR. Las Vegas, US (2016)

\bibitem{Wang2015}
Wang, Y., Wang, J., Chang, S.F.: {CamSwarm: Instantaneous Smartphone Camera
  Arrays for Collaborative Photography}.
\newblock arXiv:1507.01148  (2015)

\bibitem{wu2019detectron2}
Wu, Y., Kirillov, A., Massa, F., Lo, W.Y., Girshick, R.: Detectron2.
\newblock \url{https://github.com/facebookresearch/detectron2} (2019)

\bibitem{Yahyavi2013}
Yahyavi, A., Kemme, B.: Peer-to-peer architectures for massively multiplayer
  online games: A survey.
\newblock ACM Comput. Surv. \textbf{46}(1), 1--51 (2013)

\bibitem{Yin2019}
Yin, G., Otis, M.D., Fortin, P., Cooperstock, J.: Evaluating multimodal
  feedback for assembly tasks in a virtual environment.
\newblock In: Proc. ACM Hum.-Comput. Interact. Glasgow, UK (2019)

\bibitem{Zou2013}
Zou, D., Tan, P.: {COSLAM: Collaborative visual slam in dynamic environments}.
\newblock IEEE Trans. on Pattern Analysis and Machine Intelligence
  \textbf{35}(2), 354--366 (2013)

\end{thebibliography}

\end{document}